\documentclass[
    twocolumn,
    amsfonts,
    amsmath,
    amssymb,
    nobibnotes,
    aps,
    prd,
    nofootinbib,
    superscriptaddress
]{revtex4-2}

\usepackage{amsmath}
\usepackage{amsfonts}
\usepackage{amssymb}
\usepackage{booktabs}
\usepackage[T1]{fontenc}
\usepackage{graphicx}
\usepackage[colorlinks=true, linkcolor=blue, urlcolor=blue, citecolor=blue]{hyperref}
\usepackage{siunitx}
\usepackage{xcolor}
\usepackage{subcaption}
\usepackage{cleveref}
\usepackage[utf8]{inputenc}

\usepackage{ragged2e}

\DeclareCaptionJustification{fulljustified}{\justifying}
\usepackage{graphicx}	
\usepackage{amsmath}	
\usepackage{amssymb}	
\usepackage{siunitx}
\usepackage{xcolor}
\usepackage{orcidlink}
\usepackage{newtxtext,newtxmath}

\graphicspath{{figures/}}

\DeclareSIUnit\year{yr}
\DeclareSIUnit\parsec{pc} 

\newcommand{\depth}{\mathcal{D}^U}

\newcommand{\UIB}{
IAC3, Universitat de les Illes Balears, Cra. de Valldemossa km 7.5, 07122 Palma, Spain
}

\begin{document}
\title{Effects of Lensing by Sgr A* on Semicoherent Searches for Continuous Gravitational Waves \\ I. Signal Model and Mismatch Studies}

\author{Aditya Kumar Sharma\,\orcidlink{0000-0003-0067-346X}}
\email{aditya.sharma@uib.es}
\affiliation{\UIB}
\author{Alicia Calafat\,\orcidlink{0009-0008-7515-6305}}
\email{alicia.calafat@uib.es}
\affiliation{\UIB}
\author{Rodrigo Tenorio\,\orcidlink{0000-0002-3582-2587}}
\affiliation{\UIB}
\author{David Keitel\,\orcidlink{0000-0002-2824-626X}}
\affiliation{\UIB}
\author{Alicia M. Sintes\,\orcidlink{0000-0001-9050-7515}}
\affiliation{\UIB}

\begin{abstract}
Continuous gravitational waves (CWs) from neutron stars beyond the Galactic Center can be lensed by the supermassive black hole in Sgr~A$^*$. Since the signal lasts far longer than the lensing time delay, the two geometric-optics images overlap and interfere. Transverse source motion during the observation makes their magnifications and relative time delay evolve, producing evolving amplitude and phase modulations absent from standard CW templates. Using lensed and unlensed injections in Gaussian noise, we quantify their effect on a Hough-based semicoherent search through a semicoherent mismatch. Recovery is governed by the competition between lensing amplification and waveform modulation. Smaller impact parameters generally improve recovery through stronger magnification, while higher transverse velocities partially offset this gain through faster modulation, especially at small impact parameters and short source--lens distances, where image interference is appreciable. The response also depends on the intrinsic signal strength. Loud signals exhibit small fractional changes, with net losses in some configurations, whereas weaker signals yield broader mismatch distributions extending to more negative values, indicating greater enhancement. A population study, sampling lensing parameters from astrophysically motivated velocity and source--lens distance distributions, reproduces the strong dependence on signal strength. Differences between populations are modest, but high-velocity, short source--lens distance populations tend towards larger mismatches. Lensing effects therefore cannot be accounted for by rescaling search sensitivity with the magnification alone; reinterpreting directed Galactic Center search sensitivities and upper limits under the lensing hypothesis requires injection studies with the full lensed waveform across the lensing parameter space.
\end{abstract}

\maketitle

\section{Introduction}

Since the first direct detection of gravitational waves (GWs) from a binary black hole merger \cite{GW150914}, the LIGO, Virgo and KAGRA (LVK) network~\cite{AdvancedLIGO,AdvancedVirgo,KAGRA} has observed several hundred signals from compact binary coalescences (CBCs) \cite{GWTC5}. These observations have provided unprecedented insights into the populations and properties of black holes (BHs) and neutron stars (NSs) \cite{GWpop1}, constrained their formation and evolution, enabled stringent tests of General Relativity \cite{TGR1,TGR2,TGR3}, and established GW observations as powerful probes of fundamental physics and cosmology \cite{GWCosmo1}.

Beyond these transient CBC signals, an important target of GW searches is continuous gravitational waves (CWs): weak, nearly monochromatic signals that may be emitted by rapidly rotating, non-axisymmetric NSs and remain in the detector band throughout an observing run \cite{CW1,CW2}. The detection of CWs would provide a unique probe of NS structure and dynamics, enabling studies of the physics of dense matter, the origin of stellar deformations and the mechanisms governing the spin evolution of NSs \cite{CW1,Owen2026,Gittins2024,Haskell2022}. These signals, however, being intrinsically weak, require searches to accumulate signal power over long observation times. The resulting computational challenge prompts CW search strategies to rely on the available source information, which determines the parameter space to be searched~\cite{RilesCW,TenorioCW,WetteCW}. Targeted searches use known pulsar timing solutions \cite{O4aKnownPulsars}, directed searches use a specified sky position but unknown frequency and spin-down parameters \cite{DS1,DS2,O4aSNR,Ming2026} and all-sky searches cover unknown frequency parameters and source positions for isolated NSs \cite{O4aAllSkyIsolated,McGloughlin2026} or NSs in binary systems \cite{O4aAllSkyBinary,O3binarywidecovas}. Although no detection has thus far been reported, the resulting upper limits have already placed stringent constraints on the amplitudes of such signals and on the underlying properties of these sources.

Among the sky regions suitable for directed searches for CWs, the Galactic Center is a particularly interesting target. The Fermi Large Area Telescope has observed extended gamma-ray emission from the central region of the Galaxy \cite{FermiLAT1,FermiLAT2}. Proposed explanations for this ``GeV excess'' include dark matter particle annihilation \cite{WIMP2, WIMP3} and an unresolved population of millisecond pulsars \cite{MSGC1,MSGC2,MSGC3,MSGC4}. The latter scenario would support the presence of a substantial population of NSs in the Galactic Center, providing further motivation for directed CW searches toward this region \cite{DS1,DS2}. 

In addition to the potential abundance of CW sources, the Galactic Center also hosts Sgr~A*, a supermassive black hole with a mass of roughly $4 \times 10^6 M_\odot$ \cite{balickandbrown1974,schodel2002star,ghez2003first,gravity2019,eht2022}. Such a massive object can gravitationally lens the GW signals originating from behind the Galactic Center. Gravitational lensing is a propagation effect when radiation traverses the curved spacetime around an intervening mass \cite{schneiderlensing}, and can produce multiple images with different magnifications, arrival times and phases \cite{EzquiagaPhase}.

Although Sgr~A* provides a possible lens for Galactic sources, most GW lensing studies so far have considered cosmological CBC sources lensed by intervening galaxies or galaxy clusters \cite{Hannuksela2019,lenssearch1,lenssearch2,lenssearch3,lenssearch4}. As the catalog of GW events continues to grow with the increasing sensitivity of GW detectors, the first detection of strongly lensed GWs, likely from merging binaries, is expected in the near future \cite{Ng2018,Wierda_2021}. Hence, considerable efforts have been devoted to understanding the astrophysical and cosmological potential of lensed CBC observations (see, e.g., \cite{Goyal2021, tgrlensing2, tgrlensing3, tgrlensing4, tgrlensing5} for tests of General Relativity with lensed GWs, \cite{cosmography1, cosmography2, cosmography3, Chen2024} for precision cosmology with strongly lensed GWs, and \cite{Hannuksela:2020xor, Li2026} for accurate localization of lensed events).

For CW sources observable with ground-based detectors, previous studies have investigated the prospects for detecting lensed signals originating behind the Galactic Center \cite{NSlensprospects1}, coherent inference of lens and source properties \cite{NSlensprospects2}, the effect of microlensing by stellar-mass objects on CWs from Galactic sources \cite{NSlensprospects3}, and the potential to probe stellar interiors through gravitationally induced phase and amplitude changes in CWs during eclipses by the Sun or a source's stellar companion \cite{NSlensprospects4}. The implications of these lensing effects for signal recovery with existing CW search methods \cite{RilesCW,TenorioCW,WetteCW}, however, remain largely unexplored.

The observational signatures of lensed CWs differ in several respects from those of lensed CBCs. In searches for strongly lensed CBC signals observed by current ground-based detectors, the time delay between the images is often sufficiently long that the images appear as distinct, temporally separated events. A CW signal, by contrast, remains observable for much longer than the characteristic time delay produced by Sgr~A*, allowing the lensed images to overlap and interfere. 

Further, CBC lensing analyses commonly treat the lensing geometry as stationary over the duration of an individual signal.\footnote{There are possible exceptions, such as lensed CBCs originating in active galactic nucleus environments where the orbital motion of the binary around the central SMBH can be substantial \cite{Leong2025,Ubach2025,martens_agn}.} For CW observations spanning months to years, transverse motion can instead produce appreciable changes in the source--lens--observer geometry. These changes produce time-dependent magnifications and interference between images, introducing additional amplitude and phase modulations in the observed signal \cite{NSlensprospects1,NSlensprospects2}. 

Because standard CW templates do not account for these modulations, lensing can change the signal recovery through the combined effects of image magnifications and waveform mismatch with the unlensed template phase model. We investigate their impact on semicoherent directed searches toward the Galactic Center by modelling the interference between the two geometric-optics images produced by Sgr~A*  under the point mass model for a source undergoing transverse motion. To quantify this effect, we apply a Hough-based\cite{AliciaHough} semicoherent search with unlensed templates to lensed and unlensed signal injections. We examine how the recovered detection statistic varies with the source position relative to the lens, the source–-lens distance, the transverse velocity, and the intrinsic signal strength.

The rest of this paper is organized as follows. We present the signal model for CWs lensed by Sgr~A*, including the effects of transverse source motion, in Sec.~\ref{sec:lensed_continuous_gravitational_waves}. We describe the semicoherent search framework and detection statistic in Sec.~\ref{sub:detection_statistics}, followed by the injection and recovery setup and the definition of the semicoherent mismatch in Sec.~\ref{sub:search_setup}. In Sec.~\ref{sec:results}, we quantify changes in signal recovery for fixed lensing configurations and for lensed source populations sampled from prescribed lensing parameter distributions. Finally, in Sec.~\ref{sec:discussion}, we discuss how these results complement previous CW lensing studies and their implications for reinterpreting the sensitivities and upper limits of directed Galactic Center CW searches under the lensing hypothesis.

\section{Lensed Continuous Gravitational Waves} \label{sec:lensed_continuous_gravitational_waves}

\subsection{Continuous Gravitational-Wave Signal Model} \label{sub:cw_signal_model}

CWs are long-duration GW signals that last in the detector band over observation times spanning months to years. Potential sources of CW emission include non-axisymmetric rotating NSs~\cite{CW1,CW2}, unstable oscillation modes such as r-modes \cite{rmode,rmode2}, freely precessing NSs \cite{freeprecession1,freeprecession2,freeprecession3} and more speculative systems such as ultralight boson clouds around rotating BHs \cite{ultralightboson1, ultralightboson2, ultralightboson3}. We adopt the standard signal model for an isolated NS rotating about a principal axis with a non-axisymmetric mass quadrupole.

For such sources, the detector strain can be written as
\begin{align}
h(t)&=F_+(t;\alpha,\delta,\psi)h_+(t; f^{(k)},\alpha, \delta, h_0, \phi_0, \cos \iota ) \nonumber \\ &+ F_\times(t;\alpha,\delta,\psi)h_\times(t; f^{(k)}, \alpha, \delta, h_0,\phi_0, \cos \iota),
\end{align}
where $F_+$ and $F_\times$ are the detector antenna pattern functions, $(\alpha,\delta)$ denote the sky position of the source, $\psi$ is the polarization angle, $f^{(k)}$ denotes the $k$-th frequency spin-down term, $h_0$ is the intrinsic strain amplitude, $\phi_0$ is the phase at reference time, and $\iota$ is the inclination angle between the spin axis and the line of sight. The two GW polarizations are
\begin{equation}
h_+(t)=h_0\frac{1+\cos^2\iota}{2}\cos\Phi(t),
\quad
h_\times(t)=h_0\cos\iota\sin\Phi(t),
\end{equation}
where $\Phi(t)$ is the phase evolution at the detector. For a rotating NS with equatorial ellipticity \mbox{$\epsilon=(I_{\rm xx}-I_{\rm yy})/I_{\rm zz}$}, the intrinsic strain amplitude is given by
\begin{equation}
h_0 = \frac{4\pi^2 G}{c^4}\frac{I_{zz}\epsilon f^2}{D_{\rm S}},
\end{equation}
where $I_{zz}$ is the principal moment of inertia about the spin axis, $D_{\rm S}$ is the source distance, and $f = 2 f_{\rm rot}$ is the GW frequency. The phase evolution $\Phi(t)$ can be written as a Taylor expansion:
\begin{equation}
\Phi(t)=\phi_0+2\pi\sum_{k=0}^{s}
\frac{f^{(k)}}{(k+1)!}
\left[t_{\rm SSB}(t)-t_0\right]^{k+1},
\end{equation}
where $t_{\rm SSB}(t)$ and $t_0$ are the arrival time of the wavefront at the Solar System Barycenter (SSB) and the reference time, respectively. From here on, we write the SSB time relative to the phase reference epoch as $\tau \equiv t_{\mathrm{SSB}} - t_{0}$. Over the observation, the detector antenna patterns, $F_+$ and $F_\times$, vary with Earth's rotation. The relation between detector time and SSB arrival time also introduces Doppler modulation. For emissions from isolated NSs, the key contributions to Doppler modulation come from the rotation of Earth with the period of a sidereal day and its revolution around the SSB with the period of a year. Throughout this work, we retain only the first frequency derivative for the CW waveform. As in previous semicoherent directed searches toward the Galactic Center~\cite{DS1,DS2}, we neglect the higher-order frequency derivatives and essentially place a single template at $f^{(k)} = 0$ for $k \geq 2$.

\subsection{Strong gravitational lensing by a point mass} \label{sub:point_mass_lens}

The signal model described above does not account for additional propagation effects such as lensing due to intervening masses. If a massive object, such as Sgr A*, lies close to the line of sight of the source and the observer, GWs can be gravitationally lensed due to the spacetime curvature produced by the ``lens'' leading to multiple images arriving at the detector with different amplitudes, phases and arrival times. Due to the compact nature of BHs, unlike cosmic lenses such as galaxies and galaxy clusters, Sgr A* can be well approximated as a point mass. In this work, we therefore adopt the point-mass lens model to describe lensing by Sgr A*. Additionally, we work within the geometric-optics approximation, which is justified for the frequency bandwidth of ground-based detectors where the GW wavelength $\lambda$ is much smaller than the characteristic length scale of the lens, given by its Schwarzschild radius $R_{\rm S}$ \cite{Lensing1,Bulashenko_2022}. For example, at a reference frequency of $100\,\mathrm{Hz}$, $\lambda/R_{\rm S} \sim \mathcal{O}(10^{-4}) \ll 1$. 

\begin{figure*}[t]
\centering
\includegraphics[width=\textwidth]{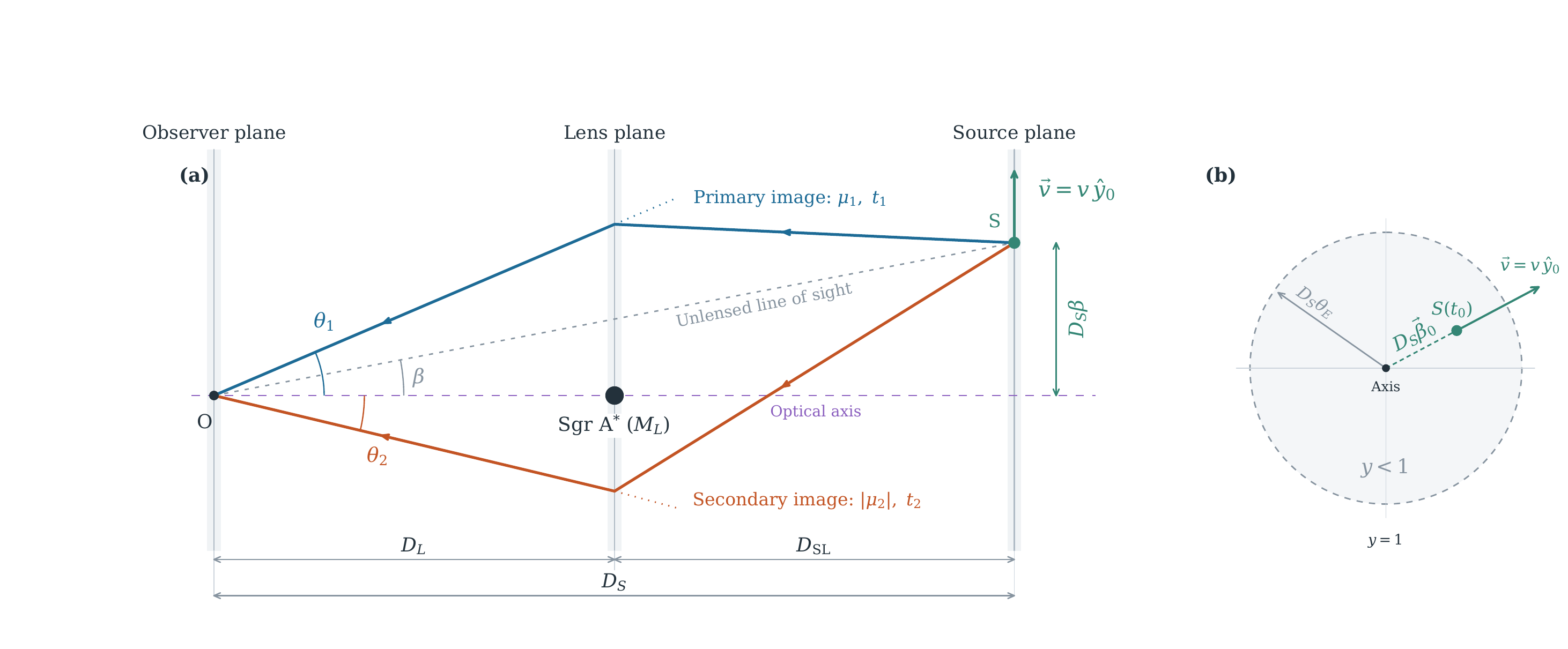}
\caption{Schematic geometry of a CW source gravitationally lensed by Sgr~A$^*$. \textbf{(a)} Observer--lens--source configuration in the thin-lens approximation. The distances $D_{\rm L}$, $D_{\rm SL}$ and $D_{\rm S}$ denote the observer--lens, source--lens and observer--source distances, respectively. The solid blue and orange paths correspond to the primary and secondary images, observed at angular positions $\theta_1$ and $\theta_2$ relative to the optical axis. The grey dashed line marks the unlensed source direction, at an angle $\beta$ to the optical axis shown in long-dashed violet. \textbf{(b)} Source-plane view showing the initial transverse offset $D_S\vec{\beta}_0$ of the source at location $S(t_0)$ on the source plane and the velocity, $v \hat{y}_0$. The dashed circle has radius $D_{\rm S}\theta_E$ and encloses the adopted strong-lensing region, $y=|\vec{\beta}|/\theta_E<1$. The angles are exaggerated for clarity in both panels. Schematic generated with the assistance of OpenAI's ChatGPT (GPT-6 Astra, Oct 2026).} 
\label{fig:lensing_schematic}
\end{figure*}

In the geometric-optics regime, gravitational lensing by a point mass produces two images of the source. The properties of these images, including their angular positions, magnifications and relative time delay, are determined by the observer--lens--source geometry and can be expressed in terms of the dimensionless source position, or impact parameter, $\vec{y}$:
\begin{equation} 
\vec{y} \equiv \frac{\vec{\beta}}{\theta_{\rm E}}.
\end{equation}
Here $\vec{\beta}$ is the angular displacement of the unlensed source from the optical axis, as seen by the observer, and
\begin{equation}
\theta_{\rm E}=\sqrt{\frac{4GM_{\rm L}}{c^2}\frac{D_{\rm SL}}{D_{\rm L}D_{\rm S}}}
\end{equation}
is the Einstein angle. Here, $M_{\rm L}$ denotes the lens mass, and $D_{\rm SL}$, $D_{\rm L}$ and $D_{\rm S}$ are the source--lens, observer--lens and observer--source distances, respectively. In the adopted Galactic geometry, $D_{\rm S} = D_{\rm L} + D_{\rm SL}$. Fig.~\ref{fig:lensing_schematic} depicts the thin-lens geometry, appropriate when the astrophysical observer–lens and lens–source distances greatly exceed the characteristic size of Sgr A*. We fix the lens mass to be the mass of Sgr A*, $4.15 \times 10^6 \, M_\odot$ \citep{SgrAstarmass}, and the distance to the lens to be the distance to the Galactic Center, $8 \, \mathrm{kpc}$ \cite{GCdistance}. 

The two images lie on opposite sides of the optical axis. Choosing the source direction from the line of sight as positive, their angular positions are
\begin{equation}
\theta_{1,2} = \frac{\theta_{\rm E}}{2}
\left( y \pm \sqrt{y^2+4} \right),
\end{equation}
and the corresponding absolute magnifications are
\begin{equation}
\left|\mu_{1,2}\right| = \frac{1}{4} \left(\frac{y}{\sqrt{y^2+4}} + \frac{\sqrt{y^2+4}}{y} \pm 2\right).
\end{equation}
The images arrive at the observer with a relative time delay,
\begin{equation} \label{eqn:timedelay1}
\Delta t =
\frac{4 G M_{\rm L}}{c^3}
\left[
\frac{y}{2}\sqrt{y^2+4}
+
\ln\left(
\frac{\sqrt{y^2+4}+y}
     {\sqrt{y^2+4}-y}
\right)
\right].
\end{equation}

The absolute magnifications of both images decrease monotonically with increasing impact parameter $y$. In the limit $y \gg 1$, the magnification of the primary image approaches unity, $|\mu_1| \rightarrow 1$, while that of the secondary image approaches zero, $|\mu_2| \rightarrow 0$. We focus on configurations with $y \leq 1$, at the reference time $t_0$, selecting sources within the Einstein radius where both images can contribute appreciably \citep{NSlensprospects1}. 

The magnification modifies the observed strain amplitude of each image such that
\begin{equation}
h_{0;1,2} = \sqrt{|\mu_{1,2}|}\,h_0,
\end{equation}
where $h_0$ is the intrinsic strain amplitude of the unlensed signal. 

The angular separation between the images is much smaller than the angular resolution of semicoherent CW searches \citep{DS2}. Therefore, we approximate both images to be at the same sky location, which we take to be the position of Sgr~A$^*$~\cite{SgrAstarmass}. This approximation is specific to our semicoherent set-up and should not be interpreted as an implication that the images are fundamentally unresolvable. For example, \citet{NSlensprospects2} showed that coherent matched-filter parameter estimation over multiple years of observation may indeed resolve these images. We provide further justification of such waveform approximations in subsequent sections.

\subsection{Lensed Continuous Gravitational-Wave Signal Model} \label{sub:lensed_cw_signal_model}

The image positions, magnifications and relative delay depend on the impact parameter, which changes under relative transverse motion and therefore modifies the observed interference pattern. We approximate this motion by a constant effective transverse velocity directed along the initial source displacement, away from the optical axis. For a general trajectory with constant transverse velocity $\vec{v}$, the direction $\hat{y}$ changes little when $|\vec{v}|T_{\rm obs}\ll D_S\theta_E y_0$, where $T_{\rm obs}$ is the observation duration. The radial velocity projection $\vec{v}\cdot\hat{y}$ then remains approximately constant, giving
\begin{equation}
    \frac{dy}{d\tau}
    = \frac{\vec{v}\cdot\hat{y}}{D_S\theta_E}
    \simeq \frac{\vec{v}\cdot\hat{y}_0}{D_S\theta_E}.
\end{equation}
In this limit, the impact parameter evolves approximately linearly. Motivated by this, we restrict to outward motion, $\vec{v}=v\hat{y}_0$ with $v>0$, so that
\begin{equation}
    \vec{y}(\tau)
    = \vec{y}_0 + \frac{\vec{v}}{D_S\theta_E}\tau,
    \label{eqn:yoft}
\end{equation}
where $\hat{y}_0=\vec{y}_0/y_0$, $y_0=|\vec{y}(0)|$ and $\Delta t_0=\Delta t(0)$. 

For a time-dependent transverse velocity, the trajectory instead satisfies
\begin{equation}
    \vec{y}(\tau)
    = \vec{y}_0
    + \frac{1}{D_S\theta_E}\int_0^\tau \vec{v}(\tau')\,d\tau',
\end{equation}
and $\frac{d\vec{y}}{d\tau}$ need not be constant.

In the hypothetical scenario of no source motion, the relative magnifications and the time delay between the two images remain constant. In a semicoherent analysis, the observed signal should therefore appear as the superposition of two monochromatic signals with a constant phase offset and a constant relative amplitude, making it indistinguishable from a typical CW signal \citep{NSlensprospects1}. However, in realistic scenarios where the source, the lens and the observer move with respect to one another, the relative time delay between the two images evolves with time. This leads to variations in the relative phase between the images, resulting in a small frequency change and consequently producing beat-like amplitude and phase modulations in the observed signal. To illustrate this effect, we will consider the phases of the two images arriving at the SSB.

\begin{subequations}\label{eq:phimoving}
\begin{align}
\Phi_1(\tau) &= \phi_0+ 2\pi\left(f\tau+\frac12\dot f\tau^2\right),
  \label{eq:phi1moving}\\
\Phi_2(\tau) &= \phi_0+2\pi\left[f\bigl(\tau-\Delta t(\tau)\bigr)
  +\frac12\dot f\bigl(\tau-\Delta t(\tau)\bigr)^2\right].
  \label{eq:phi2moving}
\end{align}
\end{subequations}
Here, the term $\Delta t$ in the phase of the second image accounts for the additional propagation delay relative to the arrival time of the first image at the observer. 

Expanding the time delay in Eq.~\ref{eqn:timedelay1} using the time-varying $y(\tau)$ in Eq.~\ref{eqn:yoft}, the arrival time delay at the SSB, to leading order, can be written as
\begin{align}
\Delta t(\tau) &\simeq \Delta t_0 + \kappa\, \tau,  \quad \textrm{where} \label{eq:timedependentdeltat1}\\
\kappa &= 2 \theta_{\rm E} \frac{D_{\rm L} }{D_{\rm SL}} \frac{v}{c}\sqrt{1+\left(\frac{y_0}{2}\right)^2}. \label{eq:kappa}
\end{align}
$\Delta t_0$ and $y_0$ are the time delay and impact parameter at the reference time. For fixed lens mass, observer--lens distance and initial impact parameter, $\kappa$ increases linearly with the transverse velocity, while increasing the source--lens distance slows the evolution of the relative time delay. Physically, a more distant source undergoes a smaller angular displacement for a given transverse velocity, leading to a slower evolution of the impact parameter and therefore the time delay. Explicitly substituting the time-dependent time delay from Eq.~\ref{eq:timedependentdeltat1} in Eq.~\ref{eq:phi2moving} and keeping only the dominant contribution, the phase of the second image has the same functional form as the first with shifted parameters, i.e.,

\begin{align}
    \Phi_2(\tau; \{f, \dot{f}\}) &\equiv \Phi_1(\tau; \{\tilde{f}, \dot{\tilde{f}}\}) \, + \Delta{\phi_0}, \label{equivalentphase} \textrm{ where} \\
    \tilde{f} &= (1 - \kappa)(f - \dot{f} \Delta t_0), \label{fshift} \\
    \dot{\tilde{f}} &= (1 - \kappa)^2 \dot{f},\label{fdotshift} \\
    \Delta{\phi_0} &= - 2 \pi f \Delta t_0  + \pi \dot{f} \Delta t_0^2. \label{phaseshift}
\end{align}

We express the resultant interfered signal at the SSB as the linear sum of the two signals with different strain amplitudes (intrinsic $h_0$ scaled with their respective magnifications $\sqrt{|\mu_{1,2}|}$), and frequency and phase offsets given by Eq.~\ref{fshift} and Eq.~\ref{phaseshift} respectively. The lensed polarizations are the sums of the two image contributions.
Defining $A_+=(1+\cos^2\iota)/2$ and $A_\times=\cos \iota$, the images' plus and cross polarizations are

\begin{subequations}\label{eq:hmoving}
\begin{align}
h_+^{\,i}(\tau)
&= A_+ h_0 \sqrt{|\mu_i(\tau)|}\cos\Phi_i(\tau).
\label{eq:hplusmoving}\\
h_\times^{\,i}(\tau)
&= A_\times h_0 \sqrt{|\mu_i(\tau)|}\sin\Phi_i(\tau).
\label{eq:hcrossmoving}
\end{align}
\end{subequations}

From here on, we will retain the intrinsic spindown in both image phases but neglect the small lensing-induced difference between their frequency derivatives. For the time-delay given in Eq.~\ref{eq:timedependentdeltat1}, the second image has $\dot{\tilde f}=(1-\kappa)^2\dot f$, giving $\Delta\dot f\simeq-2\kappa\dot f$. The lensing induced difference in frequency derivative therefore accumulates a total frequency shift of $|\Delta\dot{f}|\,T_{\rm obs} \simeq 2|\kappa\dot{f}|T_{\rm obs}$ over the full observation. This is much smaller than the leading lensing-induced frequency shift $\kappa f$ by a factor $2|\dot f|T_{\rm obs}/f$. For $\dot f=-10^{-10}\,\mathrm{Hz\,s^{-1}}$, $T_{\rm obs}=1\,\mathrm{yr}$, and $f=100$--$500\,\mathrm{Hz}$, this factor is approximately $10^{-5}$--$10^{-4}$, and hence the shift is ignored in the waveform model of the individual images. However, for fully coherent analyses, such as post-detection parameter estimation, such contributions are important to be considered \citep{NSlensprospects2}.

The resulting total plus polarization strain from the linear superposition of the two images is:

\begin{equation}
\begin{aligned}
    h_{+}^{\mathrm{tot}}(\tau)
    &= A(\tau)\cos\biggl[
        2\pi\left(f\tau+\frac{1}{2}\dot f\tau^2\right)+\phi_0 \\
    &{}+\frac{\Omega\tau}{2}
        +\frac{\varphi_{\rm rel,0}}{2}
        +\varphi_{\rm mod}(\tau)
    \biggr],
\end{aligned}
\label{htot}
\end{equation}
where:

\begin{multline}
    A(\tau) = A_+ h_0 \times \\
    \shoveleft{\sqrt{
        |\mu_1(\tau)| + |\mu_2(\tau)|
        + 2\sqrt{|\mu_1(\tau)\mu_2(\tau)|}
        \cos(\Omega\tau + \varphi_{\rm rel,0})
    },}
    \label{ampmod}
\end{multline} 
\begin{align}
    \Omega \, &= \, \frac{ - 4 \pi \, f\,  D_{\rm L} \, \theta_{\rm E} \, v}{c \, D_{\rm SL}} \sqrt{1 + \frac{y_0^2}{4}}, \label{eq:omega} \\
    \varphi_{\rm rel,0}\, &= \, - 2 \pi f \Delta t_0 , \label{eq:phirel}\\
    \varphi_{\rm mod}(\tau) &= \arctan\Biggl(\frac{\sqrt{|\mu_1(\tau)|}-\sqrt{|\mu_2(\tau)|}}{\sqrt{|\mu_1(\tau)|}+\sqrt{|\mu_2(\tau)|}}\tan\biggl(-\frac{\Omega\tau+\varphi_{\rm rel,0}}{2}\biggr)\Biggr). \label{phasemod}
\end{align}
The interference model used in our injections omits the constant $\pi/2$ Morse phase of the saddle image~\cite{Lensing1}. Including this phase would shift the interference pattern while leaving the beat frequency and image magnifications unchanged. For signals observed over many beat cycles, this shift has little effect on the overall semicoherent recovery. 

An analogous expression holds for the total cross polarization $h_\times^{\mathrm{tot}}(\tau)$, with $A_+$ replaced by $A_\times$ in Eq.~\ref{ampmod} and the outer cosine replaced by a sine in Eq.~\ref{htot}. The lensed CW signals include the amplitude modulation described by Eq.~\ref{ampmod} and the phase modulation described by Eqs.~\ref{eq:omega}--\ref{phasemod}, neither of which are accounted for in conventional searches for CW signals. 

It should be stated that we have made one further approximation when linearly summing the two images. We evaluate the detector response and Doppler modulation at a common sky position for both images. This approximation neglects the differential Doppler shift associated with their angular separation. Its effect at the semicoherent search stage is small since \mbox{$f|\vec{v}_{\rm det}\cdot(\hat{n}_2 - \hat{n}_1)|/c\ll1/T_{\rm coh}$} throughout the observation, where $T_{\mathrm{coh}}$ is the coherent integration duration of each data segment.

\begin{figure*}[t]
    \centering
    \includegraphics[width=\textwidth]{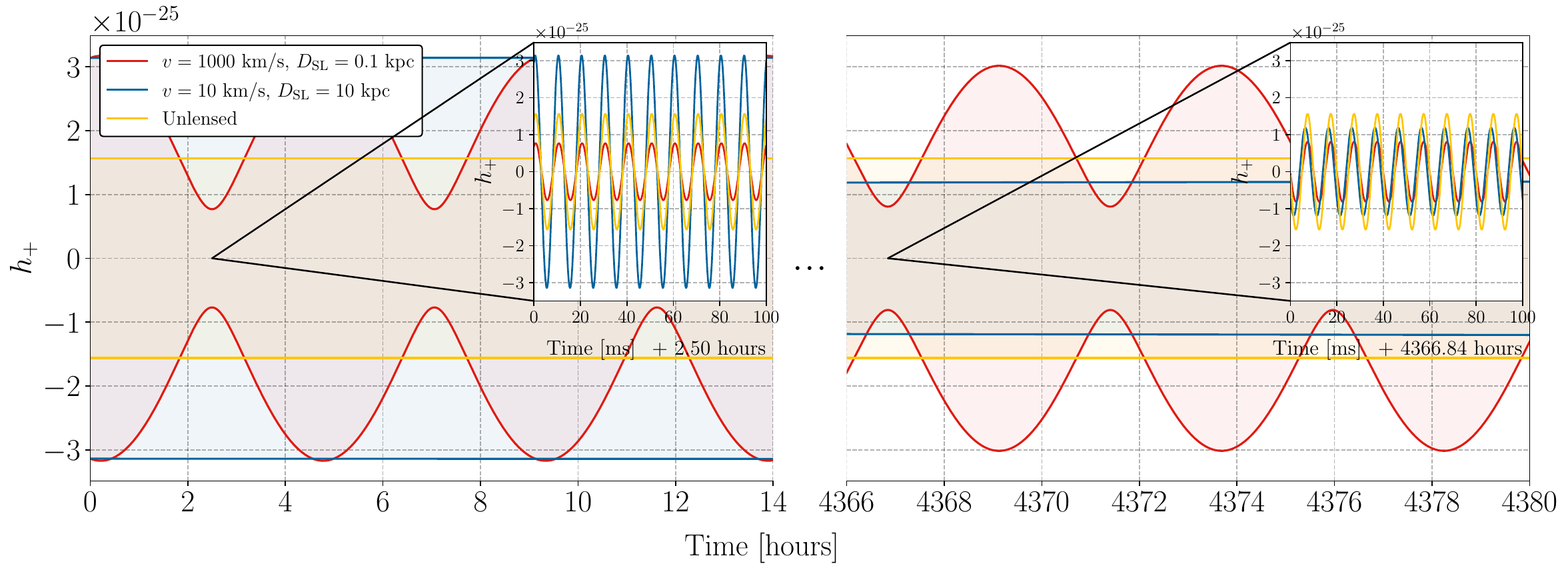}
    \caption{Example realizations of the plus polarization of a strongly lensed CW for two different source transverse velocities, compared with the corresponding unlensed signal. All examples use $f=100\,\mathrm{Hz}$, $\dot f=-10^{-10}\,\mathrm{Hz\,s^{-1}}$, $h_0=2.5\times10^{-25}$, $\cos\iota=0.5$ and $y_0=0.5$. The red signal has $v=1000\,\mathrm{km\,s^{-1}}$ and $D_{\rm SL}=0.1\,\mathrm{kpc}$, while the blue signal has $v=10\,\mathrm{km\,s^{-1}}$ and $D_{\rm SL}=10\,\mathrm{kpc}$. These extreme illustrative configurations were chosen to highlight the effects of lensing-induced amplitude and phase modulation. The main panels show the waveform envelopes and the insets show the actual oscillations. In the left panel, the slowly moving source undergoes predominantly constructive interference and its strain amplitude exceeds that of the unlensed signal. In the right panel, it undergoes predominantly destructive interference and its strain amplitude is smaller than that of the unlensed signal. Owing to its shorter beating period, the faster-moving source cycles between constructive and destructive interference more rapidly over the same time interval.}    
    \label{fig:phase_amplitude_modulations}
\end{figure*}

The phase modulation depends on both $\Omega$ and the magnification of the two images, $\mu_{^1_2}$ through Eq.~\ref{phasemod}. Here, $\Omega$ is the angular-frequency difference between the images, and $|\Omega|$ is the beat angular frequency. This modulation rate depends on the source transverse velocity and the lensing geometry through Eq.~\ref{eq:omega}. With the other parameters held fixed, faster transverse motion or a smaller source--lens distance produces more rapid beating. It is also useful to note an important distinction between lensing and Doppler modulation here. While the latter originates due to the component of the relative source-observer velocity along the line of sight, the lensing-induced modulation arises predominantly from the relative transverse motion across the lens plane. Throughout this work, $v$ represents the effective transverse velocity due to the relative motion of the source, lens and observer. Fig.~\ref{fig:phase_amplitude_modulations} depicts these characteristic beating patterns discussed above for different illustrative lens parameters. 

\section{Analysis Set-up} \label{sec:analysis}
\subsection{Semicoherent search framework} \label{sub:detection_statistics}

For strongly lensed CBC signals in the geometric-optics limit with sufficiently long time delays, the effect of lensing is straightforward to interpret in the context of matched-filter searches \cite{gwtc50methodsidentifyingcharacterizing}. Each image is recovered independently with its respective signal-to-noise ratio (SNR) rescaled by the square root of its magnification. Consequently, the impact of lensing on the search sensitivity is readily characterized through this SNR amplification. \footnote{Note that in the wave-optics regime or for overlapping geometric optics images however, diffraction and interference can introduce frequency-dependent amplitude and phase changes for CBCs. The recovered SNR then depends on both the lensing amplification and the mismatch between the lensed and the unlensed waveforms~\cite{Lensing1,overlapping1,overlapping2,overlapping3,overlapping4}.} 

The situation is fundamentally different for CWs. First, although lensing is still treated here in the geometric-optics limit, the long-lived nature of CW signals causes the multiple lensed images to interfere, producing the amplitude and phase modulations in the waveform as previously described in Sec.~\ref{sub:lensed_cw_signal_model}. Second, directed CW searches do not employ a single coherent matched-filter SNR over the entire observation time. Instead, they use semicoherent strategies to divide the entire data into shorter segments, compute coherent detection statistics over these shorter time segments and incoherently combine them, according to a prescription specific to each search method \cite{RilesCW,TenorioCW,WetteCW}. This compromise allows CW searches to be performed across wider parameter spaces for a lower computational cost than fully coherent searches, albeit at the expense of sensitivity. Thus, the impact of strong lensing cannot be inferred from a simple rescaling of the matched-filter SNR, but must instead be assessed through the response of the relevant semicoherent searches to the lensing induced features in the signal waveform.

We use a Hough-based semicoherent search that selects peaks in the power spectra of short Fourier Transforms (SFTs), and accumulates weighted number counts along frequency tracks predicted by unlensed CW signal models. The search follows the Hough construction described in \citet{AliciaHough}, with the threshold and local-maximum peak-selection criterion of \citet{PiaHough}. We implement this set-up using \texttt{FastTracks}~\cite{Fasttracks}, emulating methods used in previous directed searches towards the Galactic Center \cite{DS1,DS2}. 

The data is divided into $N$ equal, non-overlapping segments of duration $T_\textrm{SFT}$. We generate SFTs containing simulated CW signals and stationary Gaussian noise using the \texttt{lalpulsar.simulateCW} module of \texttt{LALSuite}~\cite{lalsuite}. We choose SFT durations such that, for a given signal frequency, the Doppler frequency drift caused by the Earth's rotation does not shift the signal by more than half a frequency bin during a single coherent segment, following \cite{AliciaHough}. Imposing this condition,
\begin{align}
    T_{\textrm{SFT}}(f) \, &= \, \sqrt{\frac{c \, T_e^2}{8 \, \pi^2 \, R_e \, f}}, \label{eq:tsft}
\end{align}
where $R_{\mathrm e}$ is the distance from Earth's rotation axis to the detector, corresponding to the radius of its circular trajectory due to Earth's rotation, and $T_e$ is the Earth's rotation period. $T_\textrm{SFT}$ in Eq.\ref{eq:tsft} is rounded up to an integer. 

The SFT time determines the frequency and frequency-derivative resolutions,
\begin{align}
    \delta f &= \frac{1}{T_\textrm{SFT}}, \qquad
    \delta\dot{f} = \frac{1}{T_\textrm{SFT}T_{\textrm{obs}}}.
\end{align}

For SFT $j$, let $\tilde s_{jk}$ denote the Fourier component at frequency $f_k$, and let $S_{n,j}(f_k)$ denote the one-sided noise PSD. We define the normalized power as:
\begin{align}
\rho_{jk} &= \frac{4|\tilde s_{jk}|^2}{T_{\mathrm{SFT}}S_{n,j}(f_k)}.
\end{align}
Computing $\rho_{jk}$ for every frequency bin in each SFT produces a time-frequency representation of the data, commonly referred to as the normalized spectrogram. 

For an isolated CW source located at sky position $\hat{n}$, the signal power is expected to follow a deterministic track in the normalized spectrogram given by

\begin{align}
    f_{\textrm{det}}(t; \{f, \dot{f}, \hat{n}\}) &= [f+\dot f\,(t-t_0)] \left[1+ \frac{\vec{v}_{\mathrm{det}}(t)\cdot\hat{n}}c \right], \label{track}
\end{align}
where $f$ and $\dot{f}$ are the template frequency coefficients at the reference epoch $t_0$, and $\vec{v}_{\rm det}(t)$ is the time-dependent detector velocity relative to the SSB. The GW phase parameters $\vec{\lambda} \equiv \{f,\dot{f}, \hat{n}\}$ of the signal can therefore be uniquely mapped to a track in the spectrogram. Since $\hat{n}$ is fixed to the Galactic Center, each track can actually be mapped to $\{f, \dot{f}\}$. 

Normalized power in each bin of the spectrogram is compared against a threshold power and demanded to be a local maximum, i.e., $\rho_{jk}> \rho_{\textrm{threshold}}$ and $\rho_{jk} > \rho_{j(k\pm1)}$ \cite{PiaHough}. If these conditions are met, the corresponding bin at time $t_j$ and frequency $f_k$ is assigned $n_j (f_k) = 1$, otherwise $n_j (f_k) = 0$. This digitizes the spectrogram to zeros and ones. We set $\rho_{\textrm{threshold}}$ to 5, corresponding to a false-alarm probability of $\approx0.07$ for choosing a local maximum bin due to Gaussian noise \cite{PiaHough}. 

Along a track for parameters $\vec{\lambda}=\{f,\dot{f},\hat{n}\}$, the weighted number count is defined as 
\begin{equation}
    n(\vec{\lambda}) =\sum_{j=1}^{N} w_j(\hat{n})\, n_j\!\left[f_{\rm det}(t_j;\vec{\lambda})\right],
\end{equation}
where $f_{\rm det}(t_j;\vec{\lambda})$ lies along the track given by Eq.~\ref{track}. The weight $w_j$ is the normalized sky and noise weight at time $t_j$ and is proportional to the antenna pattern average response to the $\hat{n}$ direction and inversely proportional to the average noise in the frequency band searched over at that instant, i.e. 

\begin{align}
    w_j(\hat{n}) &\propto \frac{a(t_j, \hat{n})^2 + b(t_j, \hat{n})^2}{S_{nj}}.
\end{align}
Here $a(t_j, \hat{n})$ and $b(t_j, \hat{n})$ are the antenna pattern function coefficients for sky position $\hat{n}$ at time-stamp $t_j$ \cite{JKS}, and $S_{nj}$ is the averaged single-sided power spectral density of noise across the bins in the frequency band under analysis. In this work, we work with Gaussian stationary noise. The noise weight $S_{nj}$ is therefore inconsequential and ignored. The weights satisfy \mbox{$\sum_{j=1}^{N} w_j = 1$}. 

With these normalized weights, $n(\vec{\lambda})$ goes from 0 to 1, i.e. if all bins along a chosen track exceed the power threshold and are local maxima with respect to their neighbourhood frequency bins, the recovered sky weighted number count (hereafter referred to as ``number count'') for that track is 1. This is used to define the critical ratio ($\Psi$):
\begin{align}
    \Psi(\vec{\lambda}) &= \frac{n(\vec{\lambda}) - \bar n_{\rm noise}}{\sigma},
\end{align}
where $\bar n_{\rm noise}$ and $\sigma$ are the mean and standard deviation of the number count expected from noise. The critical ratio measures the number count excess above the expected noise mean in units of the noise standard deviation. We use it as the semicoherent ranking statistic.

\subsection{Injection set-up} \label{sub:search_setup}

To quantify the impact of strong lensing on CW recovery, we compare the detection statistics of matched unlensed and lensed injections. Each matched set contains an unlensed signal and lensed counterparts with identical intrinsic phase parameters, $\vec{\lambda}\equiv\{f,\dot{f},\hat{n}\}$, amplitude parameters, $\vec{\mathcal{A}}\equiv\{h_0,\cos\iota,\psi,\phi_0\}$, and the same noise realization. Within each set, we vary only the lensing parameters,
\begin{align}
    \lambda_{\rm L} &= \{y_0,D_{\rm SL},v\}.
\end{align}
For all injections, $\hat{n}$ is fixed to the Galactic Center, with right ascension $\alpha=17^{\rm h}45^{\rm m}40^{\rm s}$ and declination $\delta=-29^{\circ}00^{\prime}28^{\prime\prime}$.

We use a nominal observation duration of $360$ days, rounded up to an integer number of complete SFTs for the LIGO Hanford detector (H1). We set the observation start and reference time $t_0$ to the start of the third LVK observing run (O3), 1 April 2019 at 15:00:00 UTC (GPS time $1238166018\,\mathrm{s}$). We generate an independent realization of stationary Gaussian noise for each set and use that same realization for its unlensed and lensed injections. 

The intrinsic signal strength is characterized by the unlensed injection depth,
\begin{align}\label{eqn:depth}
    \depth = \frac{\sqrt{S_n(f)}}{h_0},
\end{align}
where $S_n(f)$ is the one-sided power spectral density of the detector noise at the signal frequency. We consider three representative signal strengths, $\depth=\{5,20,40\} \, \textrm{Hz}^{-1/2}$, spanning signals from relatively loud to weak. Here, the limit of $\depth = 40\,\mathrm{Hz^{-1/2}}$ is not set by the sensitivity of previous directed searches \cite{DS1,DS2}, but is chosen such that the selected depths together capture the qualitative dependence of lensing effects on intrinsic signal strength. The superscript $U$ indicates that the signal depth is defined using the unlensed strain amplitude. Hence, $h_0$ denotes the strain amplitude that the source would produce at the detector in the absence of lensing. The lensed waveform is obtained by applying the additional image magnifications and time delay contribution to its unlensed counterpart signal, as in Eq.~\ref{htot}. The SFT duration varies with GW frequency according to Eq.~\ref{eq:tsft}; the values used in the analysis are listed in Table~\ref{tab:tsft-frequency}.

\begin{table}[t]
    \centering
    \caption{
        SFT durations used at the different GW frequencies considered
        in this analysis, computed using Eq.~\ref{eq:tsft}.
    }
    \label{tab:tsft-frequency}
    \begin{tabular}{cc}
        \hline\hline
        Frequency (Hz) & $T_{\rm SFT}$ (s) \\
        \hline
        100 & 7178 \\
        200 & 5076 \\
        300 & 4144 \\
        400 & 3589 \\
        500 & 3210 \\
        \hline\hline
    \end{tabular}
\end{table}

For each injection, the \texttt{FastTracks} implementation of the search described in the previous section is performed over a rectangular grid of $ \sim 1000$ templates surrounding the injected signal parameters. The grid position is randomized using independent uniform offsets spanning one template spacing in each dimension such that the injection is not, in general, located exactly on a template. The template spacings in frequency and frequency derivative are, respectively, chosen to be: 
\begin{align}
    \Delta f &= \frac{\delta f}{K_f}, \\
    \Delta\dot{f} &= \frac{\delta\dot{f}}{K_{\dot{f}}},
\end{align}
using oversampling factors $K_f=10$ and $K_{\dot{f}}=2$, as used in previous directed Galactic Center searches \citep{DS1,DS2}. 

For an unlensed injection, let $\Psi_U$ denote the maximum critical ratio and $n_U$ the corresponding maximum number count recovered across the template bank. Likewise, for the corresponding lensed injection, let $\Psi_L$ and $n_L$ denote the maximum recovered critical ratio and number count, respectively. Since the lensed and unlensed injections differ only through the lensing parameters, differences in the recovered detection statistic directly quantify the influence of strong lensing on the semicoherent search. 

We define the semicoherent mismatch between the recovery of a lensed injection with unlensed templates against the recovery of a corresponding unlensed injection as the fractional change in the recovered critical ratio,
\begin{align}
    \mathrm{mismatch}
    &= 1 - \frac{\Psi_L}{\Psi_U} \nonumber \\
    &= \frac{n_U-n_L}{n_U-\bar n_{\rm noise}},
    \label{mismatch}
\end{align}
where $\bar n_{\rm noise}$ is the mean number count expected from noise. The fractional comparison is well conditioned only when the unlensed excess $n_U-\bar n_{\rm noise}$ is sufficiently positive, a condition satisfied by all injections considered here. A positive mismatch corresponds to a reduction in the recovered critical ratio, indicating that the lensing-induced amplitude and phase modulations inhibit the recovery compared to its unlensed counterpart. Unlike the conventional notion of template mismatch, however, the mismatch defined in Eq.~\ref{mismatch} can also take negative values when the amplification due to lensing outweighs the loss due to waveform modulation, resulting in a higher recovered number count, and hence a larger critical ratio than for the corresponding unlensed injection. 

\section{Results} \label{sec:results}

We quantify the effects of lensing-induced amplification and waveform modulation on CW recovery using the semicoherent search described in Sec.~\ref{sub:detection_statistics}. We first compare representative unlensed and lensed injections to illustrate how the initial impact parameter and transverse velocity affect the recovered statistic. We then construct mismatch distributions from $100$ matched noise realizations per configuration across a fixed grid of impact parameters, transverse velocities, source--lens distances, signal depths and frequencies. Finally, we sample the lensing parameters from prescribed distributions to examine the mismatch for ensembles of lensed sources.

\subsection{Mismatch trends for fixed lensing configurations} \label{sec:illustrative}

We consider GW frequencies $100 \textrm{ to } 500 \,\mathrm{Hz}$, frequency derivative $\dot f=-10^{-10}\,\mathrm{Hz\,s^{-1}}$, and unlensed signal depths $\depth=\{5,20,40\}\,\mathrm{Hz}^{-1/2}$. The lensing parameter grid comprises $y_0=\{0.1,0.5,1.0\}$, $D_{\rm SL}=\{0.1,1.0\}\,\mathrm{kpc}$, and $v=\{100,300,500\}\,\mathrm{km\,s^{-1}}$. The range of velocity $v$ is chosen in consistency with the expected Galactic velocity dispersion \cite{schodel2009}, and the range of NS natal kick velocities \cite{Hobbs2005,Frank2017}. For each choice of $f$, $\dot f$ and $\mathcal{D}^{U}$, we generate 100 independent realizations of stationary Gaussian noise. Each realization is shared by an unlensed reference and its lensed counterparts across the lensing parameter grid. Their intrinsic phase and amplitude parameters, source orientation and template-bank placement are identical within each matched set, so differences in the recovered statistic arise only due to lensing effects. We quantify these differences using the semicoherent mismatch defined in Eq.~\ref{mismatch}. A positive mismatch indicates a reduction in the recovered critical ratio, while a negative mismatch indicates an enhancement relative to the corresponding unlensed injection. 

\begin{figure*}[t]
    \centering
    \begin{subfigure}{0.44\textwidth}
        \centering
        \includegraphics[width=\linewidth]{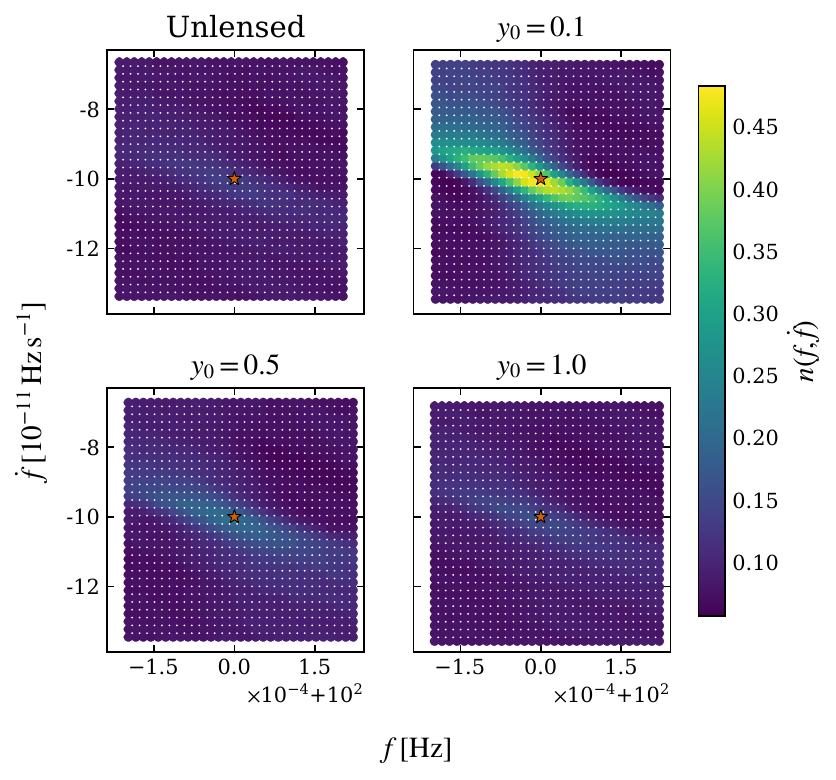}
        \caption{Number count recovered across the two-dimensional template grid of $f$ and $\dot{f}$.}
        \label{fig:example_number_count_y_a}
    \end{subfigure}
    \hfill
    \begin{subfigure}{0.48\textwidth}
        \centering
        \includegraphics[width=\linewidth]{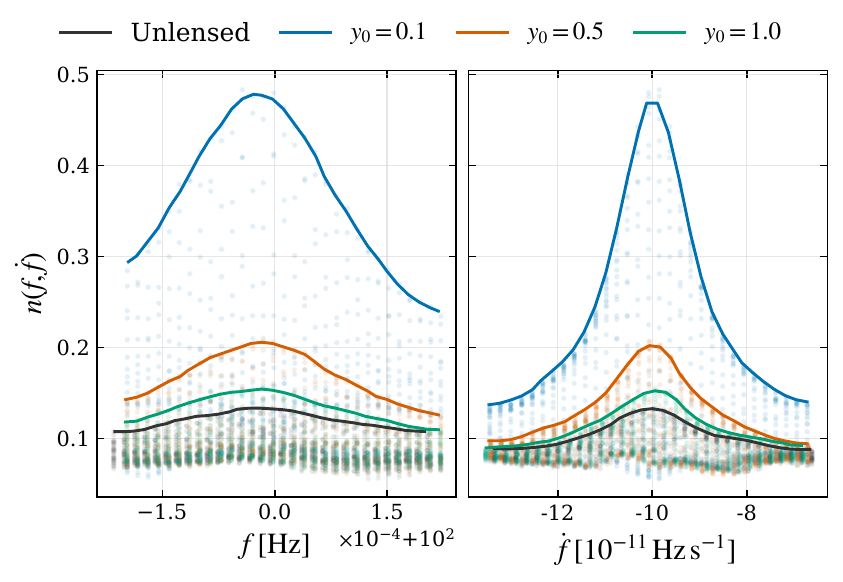}
        \caption{Number count recovered against template values of $f$ and $\dot{f}$ individually.}
        \label{fig:example_number_count_y_b}
    \end{subfigure}

    \caption{Number count recovered for an unlensed injection and lensed injections with $y_0=0.1,0.5,1.0$ at $f=100\,\mathrm{Hz}$, $\dot f=-10^{-10}\,\mathrm{Hz\,s^{-1}}$, $\mathcal{D}^{U}=40\,\mathrm{Hz}^{-1/2}$, $v=100\,\mathrm{km\,s^{-1}}$ and $D_{\rm SL}=0.1\,\mathrm{kpc}$. We use $\cos\iota = 0.5$, the polarization angle $\psi = \pi/6$, and $\phi_0 = 0$. The injections share the same noise realization, source orientation and template bank. Panel~(a) shows the count across the two-dimensional frequency--spindown grid; the stars mark the injected unlensed phase parameters. Panel~(b) displays the same template counts against $f$ and $\dot f$ separately. For this configuration, the maximum recovered count increases as $y_0$ decreases.}
    \label{fig:impact_parameter_example}
\end{figure*}

\begin{figure*}[htbp]
    \centering
    \begin{subfigure}{0.44\textwidth}
        \centering
        \includegraphics[width=\linewidth]{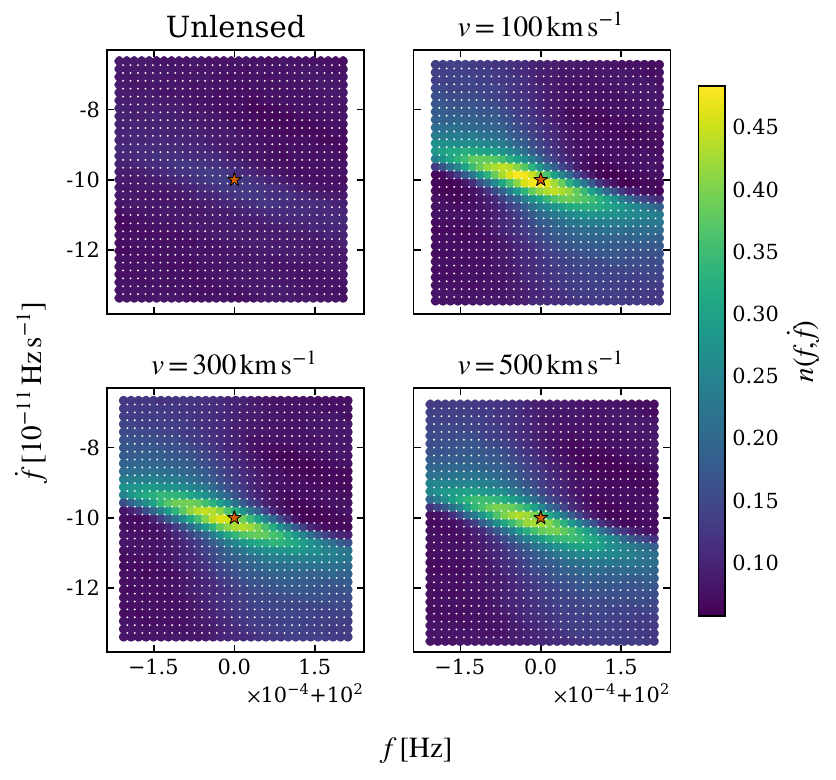}
        \caption{Number count recovered across the two-dimensional template grid of $f$ and $\dot{f}$.}
        \label{fig:example_number_count_velocity_a}
    \end{subfigure}
    \hfill
    \begin{subfigure}{0.55\textwidth}
        \centering
        \includegraphics[width=\linewidth]{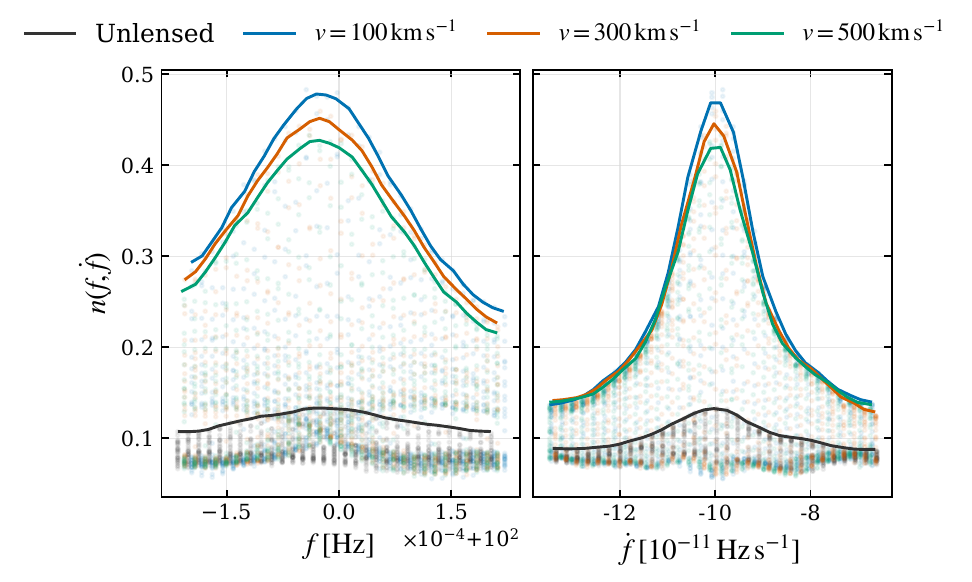}
        \caption{Number count recovered against template values of $f$ and $\dot{f}$ individually.}
        \label{fig:example_number_count_velocity_b}
    \end{subfigure}
    
    \caption{Number count recovered for an unlensed injection and lensed injections with $v=100,300,500\,\mathrm{km\,s^{-1}}$ at $f=100\,\mathrm{Hz}$, $\dot f=-10^{-10}\,\mathrm{Hz\,s^{-1}}$, $\mathcal{D}^{U}=40\,\mathrm{Hz}^{-1/2}$, $y_0=0.1$ and $D_{\rm SL}=0.1\,\mathrm{kpc}$. We use $\cos\iota = 0.5$, the polarization angle $\psi = \pi/6$, and $\phi_0 = 0$. The injections share the same noise realization, source orientation and template bank. Panel~(a) shows the count across the two-dimensional frequency--spindown grid; the stars mark the injected unlensed phase parameters. Panel~(b) displays the same template counts against $f$ and $\dot f$ separately. All three lensed injections result in a higher maximum count recovered than the unlensed reference, but the enhancement decreases as $v$ increases due to higher amplitude and phase modulation rate of the signal.}
    \label{fig:velocity_example}
\end{figure*}

Figure~\ref{fig:impact_parameter_example} presents one illustrative case of the response of the search to the impact parameter. The panels show the number count across the frequency--spindown template bank for an unlensed injection and three lensed injections with $y_0=1.0$, $0.5$, and $0.1$. The remaining parameters are fixed to $\depth=40~{\rm Hz}^{-1/2}$, $v=100~{\rm km\,s^{-1}}$, and $D_{\rm SL}=0.1~{\rm kpc}$. For the unlensed injection, the recovered number count is largest near the injected frequency and spindown. At $y_0=1$, the lensed response differs only modestly from the unlensed response, consistent with the relatively weak secondary image and modest image magnifications. Decreasing $y_0$ increases the overall magnifications of both images and, for the configuration shown in Fig.~\ref{fig:impact_parameter_example}, increases the maximum recovered number count and the enhancement is largest at $y_0=0.1$. The recovery also depends on the subsequent evolution of the magnifications and interference pattern, which we examine by varying the transverse velocity.

Figure~\ref{fig:velocity_example} shows the corresponding dependence on the source transverse velocity for fixed $y_0=0.1$, $D_{\rm SL}=0.1~{\rm kpc}$, and $\depth=40~{\rm Hz}^{-1/2}$. All three lensed injections yield a higher maximum number count than the unlensed reference in this example, but the enhancement decreases as $v$ increases from $100$ to $500\,\mathrm{km\,s^{-1}}$. This behaviour reflects the competition between magnification and modulation. At fixed $y_0$ and $D_{\rm SL}$, the image magnifications at the reference time, i.e. $\left|\mu_{1,2}(\tau=0)\right|$, are identical across the three configurations, whereas the beat angular frequency satisfies $|\Omega|\propto v$. A larger velocity therefore produces more rapid amplitude and phase evolution over the observation time. The resulting signal has faster amplitude modulation along with the power being redistributed over adjacent frequency bins and is less completely recovered by any single unlensed template. Thus, even when lensing provides a net enhancement, increasing the velocity can reduce the magnitude of that enhancement. A smaller contribution arises from the outward transverse motion adopted in Eq.~\ref{eqn:yoft}, i.e., at larger $v$, $y$ increases more rapidly with time, causing the image magnifications to decrease more during the observation.

\begin{figure*}[t]
    \centering
    \includegraphics[width=0.95\textwidth]{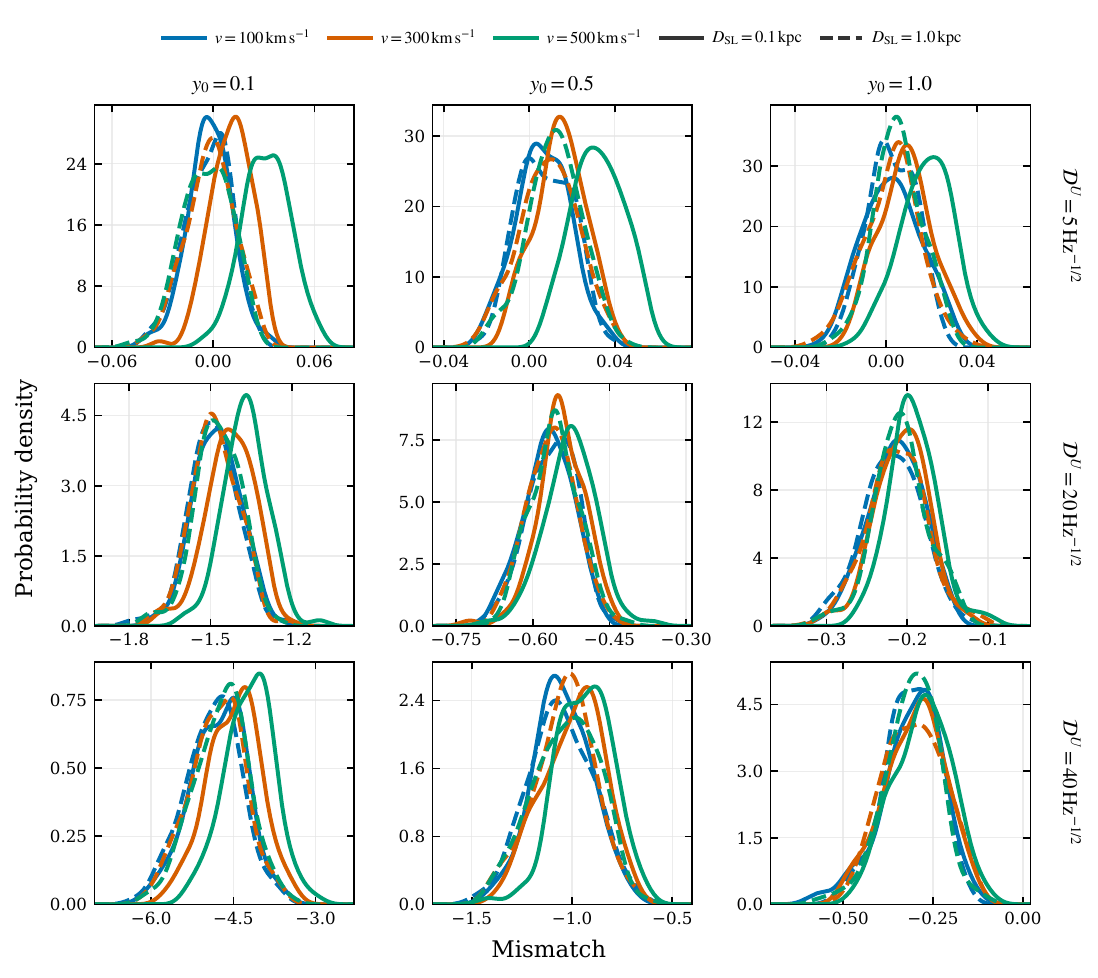}
    \caption{Gaussian kernel-density estimates of the semicoherent mismatch defined in Eq.~\ref{mismatch}, obtained from 100 matched lensed--unlensed injection pairs per lensing configuration at $f=100\,\mathrm{Hz}$ and $\dot f=-10^{-10}\,\mathrm{Hz\,s^{-1}}$. Rows correspond to unlensed depths $\mathcal{D}^{U}=5,20,40\,\mathrm{Hz}^{-1/2}$, from top to bottom, and columns to initial impact parameters $y_0=0.1,0.5,1.0$, from left to right. Blue, orange and green curves denote $v=100,300,500\,\mathrm{km\,s^{-1}}$, respectively. Solid and dashed curves denote $D_{\rm SL}=0.1$ and $1.0\,\mathrm{kpc}$, respectively. Negative mismatch indicates enhanced recovery relative to the matched unlensed reference. Increasing the unlensed depth, corresponding to weaker injections, shifts the distributions towards more negative mismatch. The velocity dependence is clearest at small $y_0$ and short source--lens distances. Axis ranges vary between panels.}
    \label{fig:mismatch_distribution_combined}
\end{figure*}

These examples illustrate how amplification and modulation jointly affect the recovery of individual injections. We now quantify this dependence across combinations of $y_0$, $v$ and $D_{\rm SL}$ at different unlensed signal depths. For each configuration, we construct mismatch distributions from matched lensed and unlensed injections using 100 independent noise realizations, with the source orientation and initial phase held fixed. Figure~\ref{fig:mismatch_distribution_combined} presents the resulting distributions at $f=100\,\mathrm{Hz}$. The parameter dependence follows from the evolution of the image magnifications and interference pattern. Smaller $y_0$ gives larger magnifications and a stronger secondary image, while $v$ and $D_{\rm SL}$ control the rate of the lensing-induced modulation. For fixed lens mass and $y_0$, Eq.~\ref{eq:omega} gives
\begin{equation}
    |\Omega|\propto f v
    \sqrt{\frac{D_{\rm L}}{D_{\rm SL}D_{\rm S}}},
    \qquad D_{\rm S}=D_{\rm L}+D_{\rm SL}.
\end{equation}

Increasing $v$ or decreasing $D_{\rm SL}$ therefore produces faster beating. These changes also affect the evolution of $y$ and hence the magnifications during the observation. The recovered mismatch depends on how the resulting signal power satisfies the peak-selection conditions along the unlensed template tracks when such effects are unaccounted for.

At fixed noise PSD, $h_0\propto(\mathcal{D}^{U})^{-1}$, so the injections at $\mathcal{D}^{U}=5\,\mathrm{Hz}^{-1/2}$ are the loudest and those at $\mathcal{D}^{U}=40\,\mathrm{Hz}^{-1/2}$ are the weakest considered here. For the intrinsically loudest injections, the mismatch distributions remain close to zero, with small positive shifts in some configurations indicating worse recovery due to the lensing effects. At $\mathcal{D}^{U}=20$ and $40\,\mathrm{Hz}^{-1/2}$, the distributions lie at increasingly negative mismatch, particularly for small $y_0$, denoting a gain in the recovery. This is consistent with the nonlinear construction of the number count. A frequency bin is selected only if its normalized power exceeds the fixed threshold and is a local maximum. The statistic is bounded above by unity, so amplification can produce only a limited gain when many bins are already selected along the recovered track, such as when the signal is intrinsically strong. For weaker signals, amplification can increase the probability of satisfying the peak-selection conditions and produce a larger fractional enhancement in the recovered statistic. The mismatch should therefore be interpreted as the response of this particular semicoherent search rather than as a waveform-only mismatch. 

At fixed $y_0$, the separation between velocity distributions is most apparent for $D_{\rm SL}=0.1\,\mathrm{kpc}$, particularly at small $y_0$ since the secondary image is sufficiently strong for interference between the two images to produce substantial waveform modulation. For $D_{\rm SL}=0.1\,\mathrm{kpc}$, the solid curves in Fig.~\ref{fig:mismatch_distribution_combined}, corresponding to different transverse velocities, generally shift towards larger mismatch as $v$ increases. At $D_{\rm SL}=1.0\,\mathrm{kpc}$, the corresponding dashed curves overlap more strongly, consistent with the slower modulation of the lensed signal at larger source--lens distances. The velocity trend is also less clear at larger $y_0$ and need not be strictly monotonic for every configuration. This is especially true for large depth and high initial impact parameter where the secondary image is too weak to contribute substantially to the signal.

Figures~\ref{fig:impact_parameter_example}, \ref{fig:velocity_example} and \ref{fig:mismatch_distribution_combined} show that, at $f=100\,\mathrm{Hz}$, lensing can either increase or decrease the statistic recovered by this semicoherent search. The sign and magnitude of the change depend on the initial impact parameter, transverse velocity, source--lens distance and unlensed signal depth. The lensed injections at $\depth=20$ and $40\,\mathrm{Hz}^{-1/2}$ show net enhancement across the configurations considered. For the loudest injections at $\depth=5\,\mathrm{Hz}^{-1/2}$, the mismatch distributions generally remain close to zero, with positive values in some configurations denoting an overall loss in comparison to their unlensed counterpart.

\begin{figure*}[t]
    \centering
    \includegraphics[width=0.95\textwidth]{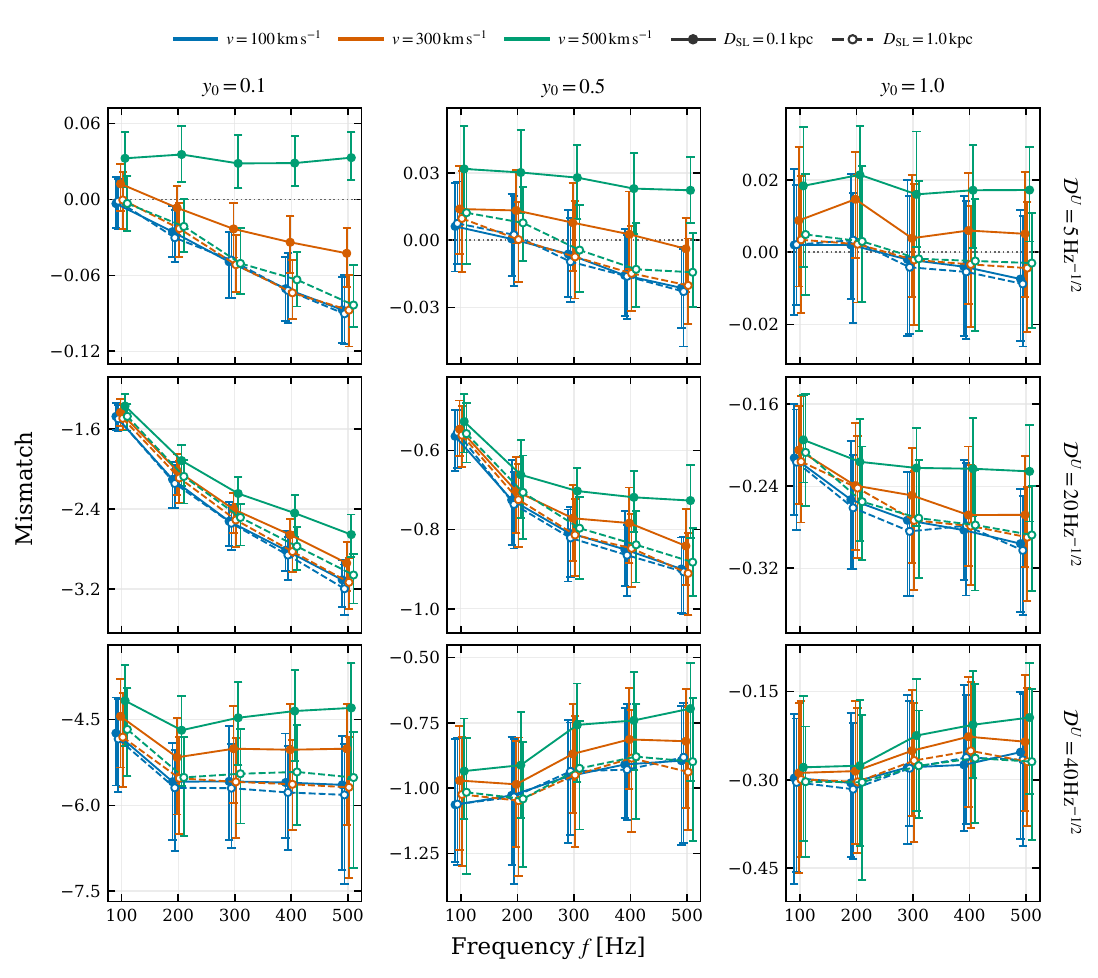}
    \caption{Median semicoherent mismatch versus injected frequency, computed from 100 matched lensed--unlensed injection pairs per frequency and lensing configuration at fixed $\dot f=-10^{-10}\,\mathrm{Hz\,s^{-1}}$. Error bars span the 5th--95th percentiles of the individual mismatch samples obtained across the realizations. Rows correspond to unlensed depths $\depth=5,20,40\,\mathrm{Hz}^{-1/2}$, from top to bottom, and columns to initial impact parameters $y_0=0.1,0.5,1.0$, from left to right. Blue, orange and green denote $v=100,300,500\,\mathrm{km\,s^{-1}}$, respectively. Solid lines with filled markers denote $D_{\rm SL}=0.1\,\mathrm{kpc}$; dashed lines with open markers denote $D_{\rm SL}=1.0\,\mathrm{kpc}$. Negative mismatch indicates enhanced recovery relative to the matched unlensed reference. Points are shifted horizontally solely for legibility.}
    \label{fig:mismatch_quantiles_all_frequency}
\end{figure*}

Figure~\ref{fig:mismatch_quantiles_all_frequency} shows that the main qualitative trends persist across the injected frequency range of $100$--$500\,\mathrm{Hz}$. The medians and associated 5th--95th percentile intervals indicate that the mismatch remains comparatively close to zero for $\mathcal{D}^{U}=5\,\mathrm{Hz}^{-1/2}$, whereas the weaker injections at $\mathcal{D}^{U}=20$ and $40\,\mathrm{Hz}^{-1/2}$ retain negative mismatch, with the largest enhancement occurring at small $y_0$. Higher transverse velocities generally shift the median towards larger mismatch, with the separation most apparent for small $y_0$ and $D_{\rm SL}=0.1\,\mathrm{kpc}$. Although the median values vary with frequency and need not evolve monotonically for every configuration, the broad dependence on lensing geometry and signal depth remains consistent with the competition between amplification and waveform modulation identified above.

Consequently, the sensitivity of a directed search toward the Galactic Center, under the lensing hypothesis, cannot be obtained by simply rescaling its unlensed strain sensitivity by the square root of the magnification. The appropriate correction is instead a distribution of search responses over the lensing parameter space. Some lensed sources that would be subthreshold in the absence of lensing may be promoted above threshold, whereas other configurations may be recovered less efficiently than their magnification alone would suggest. 

A consistent reinterpretation of an existing search would therefore require injecting the lensed signal model into the search pipeline and marginalizing the resulting detection efficiency over the relevant lensing parameters. The results presented here establish how that efficiency depends on the individual lensing and signal parameters.

\subsection{Population-level mismatch trends}
\label{subsec:population_models}

The examples above show how changes in individual lensing parameters affect the recovered semicoherent statistic. Reinterpreting CW search results under the lensing hypothesis requires marginalisation over unknown lensing parameters, in addition to the usual marginalisation over source amplitude parameters. To qualitatively illustrate how the search response depends on the adopted lensing distributions, we construct mismatch distributions using different prescriptions for the source--lens distance and transverse velocity. We use the same search set-up as described in Sec.~\ref{sub:search_setup}. For each realization, we draw $\cos\iota$ uniformly from $[-1,1]$, $\psi$ uniformly from $[-\pi/4,\pi/4]$ and $\phi_0$ uniformly from $[0,2\pi)$. The unlensed signal and all its lensed counterparts share these values given a realization.

We consider six populations by combining two transverse-velocity distributions with three source--lens distance distributions, defined later in this section. All six populations share the same initial impact-parameter distribution, restricted to the strong-lensing interval $0\leq y_0\leq1$. Assuming a uniform projected source density within the Einstein radius, we adopt
\begin{equation}
    p(y_0)=2y_0,
    \qquad 0\leq y_0\leq1.
    \label{population_y_density}
\end{equation}

To explore different velocity prescriptions within the broad range of NS transverse velocities inferred from pulsar proper-motion measurements \cite{Hobbs2005}, we adopt uniform distributions over two nonoverlapping intervals,
\begin{equation}
    50\leq v_{\rm low}\leq200\,\mathrm{km\,s^{-1}},
    \qquad
    300\leq v_{\rm high}\leq1000\,\mathrm{km\,s^{-1}}.
    \label{population_velocity_intervals}
\end{equation}

For the source--lens distance $D_{\rm SL}$, we adopt a distribution based on the Galactic radial distribution of neutron-star birth locations introduced by \citet{Paczynski1990} and subsequently used in estimates of neutron stars strongly lensed by Sgr~A$^*$ \cite{NSlensprospects1}. We refer to this as the progenitor distribution,
\begin{equation}
    p_{\rm prog}(D_{\rm SL})
    \propto
    D_{\rm SL}
    \exp\!\left(-\frac{D_{\rm SL}}{r_0}\right),
    \qquad
    0\leq D_{\rm SL}\leq15\,\mathrm{kpc},
    \label{population_progenitor_density}
\end{equation}
with $r_0=4.5\,\mathrm{kpc}$. Here, this radial dependence is adopted as a sampling prescription to sample $D_{\rm SL}$ extending up to very large values. For the two other distance distributions, we adopt the quadratic distance prior associated with the radial distribution of a uniform sphere,
\begin{equation}
    p_{\rm sph}(D_{\rm SL})
    \propto D_{\rm SL}^{2},
    \qquad
    D_{\min}\leq D_{\rm SL}\leq D_{\max}.
    \label{population_spherical_density}
\end{equation}

We distinguish these two distributions by using $(D_{\min},D_{\max})=(0.01,0.1)\,\mathrm{kpc}$ for the short-distance interval and $(D_{\min},D_{\max})=(0.1,1)\,\mathrm{kpc}$ for the larger interval.

At a fixed signal frequency of $100$ Hz, we evaluate the six populations at the three depths $\depth=5,20,40\,\mathrm{Hz}^{-1/2}$. We generate $500$ matched realizations, each containing one unlensed injection and one lensed injection from each of the six populations. Within each realization, all injections share the same stationary Gaussian noise, values of $\cos\iota$, $\psi$ and $\phi_0$, and the $f, \dot{f}$ template bank, with the sky position fixed to the Galactic Center. The lensing parameters are sampled according to the corresponding population prescriptions. We reuse this same set of $500$ realizations of the noise, source and lensing parameters, and template banks. This construction reduces variations unrelated to the population prescriptions when comparing the resulting mismatch distributions.

\begin{figure}[t]
    \centering
    \includegraphics[
        width=\columnwidth
    ]{
        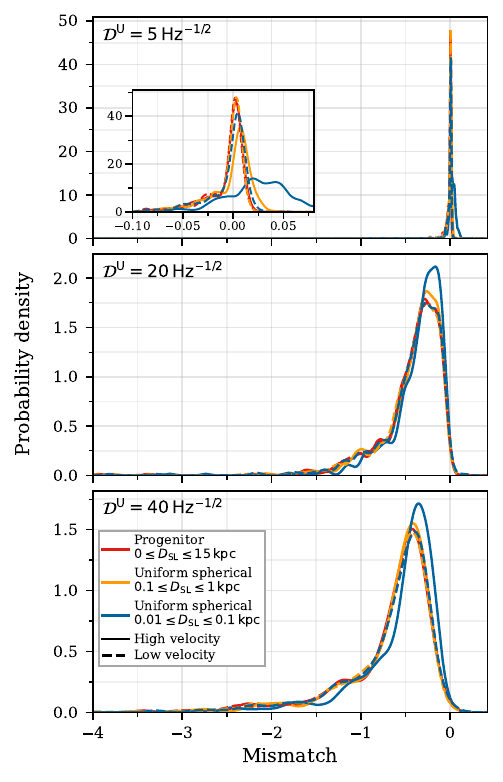
    }
    \caption{
        Gaussian kernel-density estimates of the mismatch distributions for the six populations at $f=100\,\mathrm{Hz}$ and $\dot f=-10^{-10}\,\mathrm{Hz\,s^{-1}}$. We draw $\cos\iota$ uniformly from $[-1,1]$, $\psi$ uniformly from $[-\pi/4,\pi/4]$ and $\phi_0$ uniformly from $[0,2\pi)$. Each curve is constructed from the $500$ mismatch values obtained for one population at the indicated depth. Red denotes the progenitor distance prescription over $0\leq D_{\rm SL}\leq15\,\mathrm{kpc}$, orange the uniform spherical prescription over $0.1\leq D_{\rm SL}\leq1\,\mathrm{kpc}$ and blue the uniform spherical prescription over $0.01\leq D_{\rm SL}\leq0.1\,\mathrm{kpc}$. Solid curves correspond to the high-velocity prescription and dashed curves to the low-velocity prescription. The inset enlarges the region around zero mismatch at $\depth=5\,\mathrm{Hz}^{-1/2}$. Negative mismatch indicates that the lensed injection yields a larger maximum number count than its corresponding unlensed injection.
    }
    \label{fig:population_mismatch}
\end{figure}

For each population and depth, the recovered semicoherent statistics of the lensed injections are compared with those of their corresponding unlensed references using the mismatch defined in Eq.~\ref{mismatch}. Figure~\ref{fig:population_mismatch} shows Gaussian kernel-density estimates of the resulting mismatch distributions at $f=100$ Hz.

The dominant trend in Fig.~\ref{fig:population_mismatch} is the dependence on signal depth, while the differences among the six population prescriptions at fixed depth are comparatively modest. At $\depth=5\,\mathrm{Hz}^{-1/2}$, all six distributions are narrowly concentrated around zero mismatch and largely overlap. The high-velocity short-distance population shows the most noticeable departure, with a somewhat broader distribution extending toward larger mismatch values. This indicates that lensing provides less enhancement of the recovered semicoherent statistic compared with the corresponding unlensed injections, with positive mismatch values indicating a net reduction in the recovered statistic relative to the unlensed case.

At $\depth=20\,\mathrm{Hz}^{-1/2}$, the distributions become substantially broader and extend farther towards negative mismatch. Differences among the population prescriptions remain relatively small compared with the overall width of the distributions, although the high-velocity short-distance population is slightly shifted towards larger mismatch than the other high-velocity populations.

At $\depth=40\,\mathrm{Hz}^{-1/2}$, the distributions are broader still and extend farther towards negative mismatch. The same qualitative tendency is visible among the high-velocity populations, with the short-distance prescription lying at somewhat larger mismatch, while the low-velocity distributions remain strongly overlapping.

These qualitative trends are consistent with the lensing behaviour described earlier. Higher transverse velocities and shorter source-lens distances lead to a faster evolution of the lensing geometry and therefore to stronger time-dependent modulation of the signal. Such modulation can reduce the agreement with the unlensed template tracks, while lensing amplification can instead increase the recovered semicoherent statistic. The increasingly negative tail at larger signal depth is consistent with the growing relative importance of amplification for weaker signals. The population study therefore shows the same qualitative competition between amplification and waveform modulation found in the fixed-parameter analysis, although the differences between the individual population prescriptions are comparatively subtle. 

The impact-parameter prior also contributes to this modest separation. Since $p(y_0)=2y_0$, the sampling favours larger initial impact parameters, for which the secondary image is weaker relative to the primary image and the interference modulation is less pronounced. The population distributions therefore give less weight to the small-$y_0$ configurations where differences between velocity and distance prescriptions are most evident.

These comparisons are conditional on the adopted distance and velocity priors. The prescriptions were chosen as representative distributions to distinguish low and high transverse velocities, and short and long source--lens distances. The resulting ensembles are not an empirically calibrated model of the strongly lensed NS population. Other sampling prescriptions may change the quantitative mismatch distributions and the separation between populations.

\section{Summary and Discussion}
\label{sec:discussion}

In this paper, we studied how strong gravitational lensing by Sgr~A$^*$ can affect the recovery of CWs from non-axisymmetric rotating neutron stars from beyond the Galactic Center. In the geometric-optics regime, the two lensed images overlap and interfere over the observing time. As the lensing geometry evolves, the image magnifications and relative time delay become time dependent, producing long-lived amplitude and phase modulations that are not captured by standard unlensed CW templates, as discussed in Sec.~\ref{sec:lensed_continuous_gravitational_waves}. 

This study complements previous studies of strong lensing of CWs that address different stages of the observational problem. Basak et al.~\cite{NSlensprospects1} estimated the number and detectability of NSs strongly lensed by Sgr~A$^*$ under astrophysically motivated Galactic source distributions. Their work established the potential lensed population of NSs and the prospects for their detection, while treating detectability primarily in terms of the lensing magnification and expected signal strength.~\citet{NSlensprospects2} considered the subsequent inference problem. By assuming a lensed CW signal can be identified, they showed how coherent matched-filtering and parameter estimation methods could distinguish the lensing hypothesis, potentially resolve the individual images, and therefore probe the properties and environment of Sgr~A$^*$. Our study complements these works on strong lensing by examining lensing in the context of realistic semicoherent CW searches. It shows how lensing affects the recovery of CW signals using standard unlensed templates and highlights its practical implications for searches directed towards the Galactic Center. 

Although the Galactic Center is a particularly promising target for lensed CWs due to the presence of Sgr~A$^*$, CW lensing is not restricted to this region. Previous work by ~\citet{NSlensprospects3} investigated CW microlensing by stellar-mass lenses in the wave-optics regime. In that scenario, the relative lens motion produces a transient magnification. These observational features were studied through the corresponding time-dependent increase in signal-to-noise ratio using a semicoherent $\mathcal{F}$-statistic search sensitive to such short-duration features \cite{JKS,FstatPrix}. In contrast, lensing by Sgr~A$^*$ produces a longer-lasting lensing signature in the geometric-optics regime, as presented in our study. 

In our work, the fixed lensing configuration study in Sec.~\ref{sec:illustrative} isolates the effects of the individual lensing and signal parameters. The initial impact parameter primarily determines the image magnifications and the relative strength of the secondary image. Smaller values of $y_0$ therefore generally produce more negative mismatch values. This effect is, however, not determined by magnification alone. The transverse velocity and source--lens distance control the rate at which the lensing geometry evolves, with the characteristic modulation rate scaling approximately as \mbox{$fv\sqrt{\left( D_{\rm L}/D_{\rm SL}D_{\rm S}\right)}$}. Increasing $v$, or decreasing $D_{\rm SL}$, consequently produces more rapid amplitude and phase modulation. Because this evolution is not described by the standard unlensed template tracks, it reduces the fraction of the lensed signal recovered by a single template and shifts the mismatch towards larger values. The velocity dependence is strongest for small $D_{\rm SL}$ and small $y_0$, where the secondary image remains sufficiently strong for two-image interference to produce appreciable modulation; it becomes progressively weaker as either $D_{\rm SL}$ or $y_0$ increases. 

These results also show a strong dependence on intrinsic signal strength. For the loudest injections, the unlensed signal already produces a large number of above-threshold time--frequency bins, so lensing causes only a comparatively small fractional change in the recovered statistic. For weaker injections, a larger fraction of the signal power lies near the peak selection threshold, and lensing amplification can promote many additional bins above it. The mismatch distributions therefore broaden and extend towards more negative values as the signal depth increases. This trend follows from the nonlinear construction of the number-count statistic and implies that the mismatch should be interpreted as the response of the type of semicoherent search considered here, rather than as a waveform-only measure usually discussed in the context of lensed CBCs. 

Taken together, the injection studies with fixed lensing and source parameters demonstrate a competition between amplification, which tends to improve recovery, and time-dependent modulation, which can partially offset that improvement by redistributing power away from the unlensed template track.  

The population study in Sec.~\ref{subsec:population_models} extends the fixed-parameter analysis to six populations combining two transverse-velocity distributions with three source-lens-distance prescriptions. The dominant trend is the dependence on signal depth. At $\depth=5\,\mathrm{Hz}^{-1/2}$, the mismatch distributions are narrow and concentrated close to zero, whereas at $\depth=20$ and $40\,\mathrm{Hz}^{-1/2}$ they become progressively broader and extend farther towards negative mismatch. Since negative mismatch corresponds to a larger recovered semicoherent statistic for the lensed injection than for its unlensed reference, this behaviour is consistent with lensing amplification becoming relatively more important for weaker signals, as discovered also in Sec.~\ref{sec:illustrative}. Differences among the population prescriptions at fixed depth are comparatively modest. The most visible departure is associated with the high-velocity short-distance population, whose mismatch distribution is shifted somewhat towards larger values relative to the other high-velocity populations. The population results therefore reproduce the same competition identified in the fixed-parameter study, i.e., lensing amplification can increase the recovered semicoherent statistic and drive the mismatch towards negative values, while time-dependent waveform modulation can reduce the agreement with the unlensed template tracks and shift the mismatch towards larger values. The balance between these effects depends on the signal strength and lensing geometry. 

Our results establish that the effect of lensing on a CW search cannot be described accurately by simply rescaling the strain upper limits according to the lensing magnification. Interpreting search results under the lensing hypothesis therefore requires the full lensed signal model to be included in injection-based sensitivity studies, which we leave to future work. 

The results presented in this work are not unique to the current LVK network of detectors. Since the analysis characterises the signal recovery in terms of the signal strength relative to the noise PSD through the injection depth $\depth$, the qualitative dependence is expected to hold for the next generation of ground-based detectors, like the Einstein Telescope \cite{ET1,ET2} and Cosmic Explorer \cite{CE1}, provided a similar semicoherent search set-up is used. Beyond ground-based detectors, analogous trends may also persist for long duration lensed sources for space-based Laser Interferometer Space Antenna (LISA) \cite{LISA1}, where semicoherent search methods are under consideration \cite{LISASemicoherent1,LISASemicoherent2}. However, for lensed sources in LISA, the relevant lensing geometry and signal frequencies will differ and a dedicated study would be needed to quantify the corresponding effects. 

\section*{Acknowledgements}
We would like to thank Ornella Piccinni and Ajith Parameswaran for useful discussions. We thank Sreekanth Harikumar for their careful reading of the manuscript and useful suggestions. This work was supported by the Universitat de les Illes Balears (UIB) with funds from the Programa de Foment de la Recerca i la Innovació de la UIB 2024-2026 (supported by the yearly plan of the Tourist Stay Tax ITS2023-086); the Spanish Agencia Estatal de Investigación grants PID2022-138626NB-I00, RED2024-153978-E, RED2024-153735-E, funded by MICIU/AEI/10.13039/501100011033 and the ERDF/EU; the Comunitat Autònoma de les Illes Balears through an Ajut per a projectes de recerca científica i tecnològica (PRD2025\_100) and through the Conselleria d'Educació i Universitats with funds from the European Union - European Regional Development Fund (ERDF) (SINCO2022/18146 - Plataforma HiTech-IAC3-BIO); and by COST action SCALES CA24139, supported by COST (European Cooperation in Science and Technology). The authors are grateful for computational resources provided by the LIGO Laboratory and supported by National Science Foundation Grants PHY-0757058 and PHY-0823459. Some of the computations were performed with the aid of the Alice Computing Cluster at ICTS-TIFR. This document has been assigned LIGO document number LIGO-P2600586. 

\bibliographystyle{apsrev4-2-short}
\bibliography{references}

@article{Paczynski1990,
author = {Paczynski, B.},
title = {{A Test of the Galactic Origin of Gamma-Ray Bursts}},
journal = {The Astrophysical Journal},
volume = {348},
pages = {485--494},
year = {1990},
doi = {10.1086/168257}
}

@article{Hobbs2005,
author = {Hobbs, G. and Lorimer, D. R. and Lyne, A. G. and Kramer, M.},
title = {{A statistical study of 233 pulsar proper motions}},
journal = {Monthly Notices of the Royal Astronomical Society},
volume = {360},
number = {3},
pages = {974--992},
year = {2005},
doi = {10.1111/j.1365-2966.2005.09087.x}
}

@ARTICLE{GW150914,
       author = {{The LIGO Scientific Collaboration} and {the Virgo Collaboration} and {Abbott}, B.~P. and {Abbott}, R. and {Abbott}, T.~D. and {Abernathy}, M.~R. and {Acernese}, F. and {Ackley}, K. and {Adams}, C. and {Adams}, T. and {Addesso}, P. and {Adhikari}, R.~X. and {Adya}, V.~B. and {Affeldt}, C. and {Agathos}, M. and {Agatsuma}, K. and {Aggarwal}, N. and {Aguiar}, O.~D. and {Aiello}, L. and {Ain}, A. and {Ajith}, P. and {Allen}, B. and {Allocca}, A. and {Altin}, P.~A. and {Anderson}, S.~B. and {Anderson}, W.~G. and {Arai}, K. and {Arain}, M.~A. and {Araya}, M.~C. and {Arceneaux}, C.~C. and {Areeda}, J.~S. and {Arnaud}, N. and {Arun}, K.~G. and {Ascenzi}, S. and {Ashton}, G. and {Ast}, M. and {Aston}, S.~M. and {Astone}, P. and {Aufmuth}, P. and {Aulbert}, C. and {Babak}, S. and {Bacon}, P. and {Bader}, M.~K.~M. and {Baker}, P.~T. and {Baldaccini}, F. and {Ballardin}, G. and {Ballmer}, S.~W. and {Barayoga}, J.~C. and {Barclay}, S.~E. and {Barish}, B.~C. and {Barker}, D. and {Barone}, F. and {Barr}, B. and {Barsotti}, L. and {Barsuglia}, M. and {Barta}, D. and {Bartlett}, J. and {Barton}, M.~A. and {Bartos}, I. and {Bassiri}, R. and {Basti}, A. and {Batch}, J.~C. and {Baune}, C. and {Bavigadda}, V. and {Bazzan}, M. and {Behnke}, B. and {Bejger}, M. and {Belczynski}, C. and {Bell}, A.~S. and {Bell}, C.~J. and {Berger}, B.~K. and {Bergman}, J. and {Bergmann}, G. and {Berry}, C.~P.~L. and {Bersanetti}, D. and {Bertolini}, A. and {Betzwieser}, J. and {Bhagwat}, S. and {Bhandare}, R. and {Bilenko}, I.~A. and {Billingsley}, G. and {Birch}, J. and {Birney}, I.~A. and {Birnholtz}, O. and {Biscans}, S. and {Bisht}, A. and {Bitossi}, M. and {Biwer}, C. and {Bizouard}, M.~A. and {Blackburn}, J.~K. and {Blair}, C.~D. and {Blair}, D.~G. and {Blair}, R.~M. and {Bloemen}, S. and {Bock}, O. and {Bodiya}, T.~P. and {Boer}, M. and {Bogaert}, G. and {Bogan}, C. and {Bohe}, A. and {Bojtos}, P. and {Bond}, C. and {Bondu}, F. and {Bonnand}, R. and {Boom}, B.~A. and {Bork}, R. and {Boschi}, V. and {Bose}, S. and {Bouffanais}, Y. and {Bozzi}, A. and {Bradaschia}, C. and {Brady}, P.~R. and {Braginsky}, V.~B. and {Branchesi}, M. and {Brau}, J.~E. and {Briant}, T. and {Brillet}, A. and {Brinkmann}, M. and {Brisson}, V. and {Brockill}, P. and {Brooks}, A.~F. and {Brown}, D.~A. and {Brown}, D.~D. and {Brown}, N.~M. and {Buchanan}, C.~C. and {Buikema}, A. and {Bulik}, T. and {Bulten}, H.~J. and {Buonanno}, A. and {Buskulic}, D. and {Buy}, C. and {Byer}, R.~L. and {Cabero}, M. and {Cadonati}, L. and {Cagnoli}, G. and {Cahillane}, C. and {Bustillo}, J. Calder{\'o}n and {Callister}, T. and {Calloni}, E. and {Camp}, J.~B. and {Cannon}, K.~C. and {Cao}, J. and {Capano}, C.~D. and {Capocasa}, E. and {Carbognani}, F. and {Caride}, S. and {Diaz}, J. Casanueva and {Casentini}, C. and {Caudill}, S. and {Cavagli{\`a}}, M. and {Cavalier}, F. and {Cavalieri}, R. and {Cella}, G. and {Cepeda}, C.~B. and {Baiardi}, L. Cerboni and {Cerretani}, G. and {Cesarini}, E. and {Chakraborty}, R. and {Chalermsongsak}, T. and {Chamberlin}, S.~J. and {Chan}, M. and {Chao}, S. and {Charlton}, P. and {Chassande-Mottin}, E. and {Chen}, H.~Y. and {Chen}, Y. and {Cheng}, C. and {Chincarini}, A. and {Chiummo}, A. and {Cho}, H.~S. and {Cho}, M. and {Chow}, J.~H. and {Christensen}, N. and {Chu}, Q. and {Chua}, S. and {Chung}, S. and {Ciani}, G. and {Clara}, F. and {Clark}, J.~A. and {Cleva}, F. and {Coccia}, E. and {Cohadon}, P.-F. and {Colla}, A. and {Collette}, C.~G. and {Cominsky}, L. and {Constancio}, M. and {Conte}, A. and {Conti}, L. and {Cook}, D. and {Corbitt}, T.~R. and {Cornish}, N. and {Corsi}, A. and {Cortese}, S. and {Costa}, C.~A. and {Coughlin}, M.~W. and {Coughlin}, S.~B. and {Coulon}, J.-P. and {Countryman}, S.~T. and {Couvares}, P. and {Cowan}, E.~E. and {Coward}, D.~M. and {Cowart}, M.~J.},
        title = "{Observation of Gravitational Waves from a Binary Black Hole Merger}",
      journal = {Physical Review Letters},
         year = 2016,
        month = feb,
       volume = {116},
       number = {6},
          eid = {061102},
        pages = {061102},
          doi = {10.1103/PhysRevLett.116.061102},
archivePrefix = {arXiv},
       eprint = {1602.03837},
 primaryClass = {gr-qc},
       adsurl = {https://ui.adsabs.harvard.edu/abs/2016PhRvL.116f1102A}
}

@ARTICLE{GWTC5,
       author = {{The LIGO Scientific Collaboration} and {the Virgo Collaboration} and {the KAGRA Collaboration} and {Abac}, A.~G. and {Abe}, A. and {Abouelfettouh}, I. and {Acernese}, F. and {Ackley}, K. and {Adam}, A. and {Adhicary}, S. and {Adhikari}, D. and {Adhikari}, R.~X. and {Adkins}, V.~K. and {Afroz}, S. and {Agapito}, A. and {Agarwal}, D. and {Agathos}, M. and {Aggarwal}, N. and {Aggarwal}, S. and {Aguiar}, O.~D. and {Ahrend}, I.-L. and {Aiello}, L. and {Ain}, A. and {Ajith}, P. and {Akutsu}, T. and {Albers}, L. and {Ali}, W. and {Al-Kershi}, S. and {Allene}, C. and {Allocca}, A. and {Al-Shammari}, S. and {Alvarez}, J.~A. and {Alvarez-Lopez}, S. and {Amar}, W. and {Amarasinghe}, O. and {Amato}, A. and {Amicucci}, F. and {Amra}, C. and {Anand}, A.~B. and {Anand}, C. and {Ananyeva}, A. and {Anderson}, S.~B. and {Anderson}, W.~G. and {Andia}, M. and {Ando}, M. and {Andrade-Oliveira}, F. and {Andr{\'e}s-Carcasona}, M. and {Andrey}, J.~L. and {Andri{\'c}}, T. and {Anglin}, J. and {Anna}, J. and {Antelis}, J.~M. and {Antier}, S. and {Aoki}, T. and {Aoumi}, M. and {Appavuravther}, E.~Z. and {Appelt}, E.~A. and {Appert}, S. and {Apple}, S.~K. and {Arai}, K. and {Araya}, A. and {Araya}, M.~C. and {Arca Sedda}, M. and {Arciprete}, F. and {Areeda}, J.~S. and {Aritomi}, N. and {Armato}, F. and {Armstrong}, S. and {Arnaud}, N. and {Arogeti}, M. and {Aronson}, S.~M. and {Ashton}, G. and {Aso}, Y. and {Asprea}, L. and {Assiduo}, M. and {Assis de Souza Melo}, S. and {Aston}, S.~M. and {Astone}, P. and {Aswathi}, P.~S. and {Attadio}, F. and {Aubin}, F. and {AultONeal}, K. and {Avallone}, G. and {Avdeev}, N. and {Avila}, E.~A. and {Babak}, S. and {Badger}, C. and {Bae}, S. and {Bagnasco}, S. and {Baimukhametova}, S. and {Baiotti}, L. and {Baka}, T. and {Baker}, K.~A. and {Baker}, T. and {Balbi}, G. and {Baldi}, G. and {Baldicchi}, N. and {Ball}, M. and {Ballardin}, G. and {Ballelli}, M. and {Ballmer}, S.~W. and {Banagiri}, S. and {Banerjee}, B. and {Bankar}, D. and {Baptiste}, T.~M. and {Baral}, P. and {Baratti}, M. and {Barayoga}, J.~C. and {Baric}, K. and {Barish}, B.~C. and {Barker}, D. and {Barman}, N. and {Barone}, F. and {Barr}, B. and {Barrios}, M. and {Barsotti}, L. and {Barsuglia}, M. and {Barta}, D. and {Barton}, M.~A. and {Bartos}, I. and {Basalaev}, A. and {Bassiri}, R. and {Basti}, A. and {Bawaj}, M. and {Bayley}, J.~C. and {Baylor}, A.~C. and {Baynard}, II, P.~A. and {Bazzan}, M. and {Bedakihale}, V.~M. and {Beirnaert}, F. and {Bejger}, M. and {Bell}, A.~S. and {Bellani}, C. and {Bellie}, D.~S. and {Beltran-Martinez}, D. and {Benedetti}, E. and {Benoit}, W. and {Bentara}, I. and {Ben Yaala}, M. and {Bera}, S. and {Bergamin}, F. and {Berger}, B.~K. and {Beroiz}, M. and {Berry}, C.~P.~L. and {Berry}, I. and {Bersanetti}, D. and {Bertheas}, T. and {Bertolini}, A. and {Betzwieser}, J. and {Beveridge}, D. and {Bevins}, N. and {Bezerra-Sobrinho}, J. and {Bhandare}, R. and {Bhatt}, R. and {Bhattacharjee}, A. and {Bhattacharjee}, D. and {Bhattacharyya}, S. and {Bhaumik}, S. and {Biancalana}, V. and {Bianchi}, F. and {Bilenko}, I.~A. and {Bilicki}, M. and {Billingsley}, G. and {Binetti}, A. and {Bini}, S. and {Biot}, S. and {Birnholtz}, O. and {Biscoveanu}, S. and {Bisht}, A. and {Bitossi}, M. and {Bizouard}, M.-A. and {Blaber}, S. and {Blackburn}, J.~K. and {Blagg}, L.~A. and {Blair}, C.~D. and {Blair}, D.~G. and {Bloch}, M. and {Bode}, N. and {Boettner}, N. and {Bogdan}, P. and {Boileau}, G. and {Boldrini}, M. and {Bolingbroke}, G.~N. and {Bonavena}, L.~D. and {Bonhomme}, V.~A. and {Bonilla}, E. and {Bonilla}, M.~S. and {Bonino}, A. and {Bonnand}, R. and {Borchers}, A. and {Borghi}, N. and {Boschi}, V. and {Bose}, S. and {Bossilkov}, V. and {Bothra}, Y. and {Boudon}, A. and {Boybeyi}, T.~D. and {Boyle}, M. and {Bozzi}, A. and {Bradaschia}, C.},
        title = "{GWTC-5.0: Observations from the Second Part of the Fourth LIGO-Virgo-KAGRA Observing Run and Updates to the Gravitational-Wave Transient Catalog}",
      journal = {arXiv e-prints},
         year = 2026,
        month = may,
          eid = {arXiv:2605.27225},
        pages = {arXiv:2605.27225},
          doi = {10.48550/arXiv.2605.27225},
archivePrefix = {arXiv},
       eprint = {2605.27225},
 primaryClass = {gr-qc},
       adsurl = {https://ui.adsabs.harvard.edu/abs/2026arXiv260527225T}
}

@ARTICLE{AdvancedLIGO,
       author = {{The LIGO Scientific Collaboration} and {Aasi}, J. and {Abbott}, B.~P. and {Abbott}, R. and {Abbott}, T. and {Abernathy}, M.~R. and {Ackley}, K. and {Adams}, C. and {Adams}, T. and {Addesso}, P. and {Adhikari}, R.~X. and {Adya}, V. and {Affeldt}, C. and {Aggarwal}, N. and {Aguiar}, O.~D. and {Ain}, A. and {Ajith}, P. and {Alemic}, A. and {Allen}, B. and {Amariutei}, D. and {Anderson}, S.~B. and {Anderson}, W.~G. and {Arai}, K. and {Araya}, M.~C. and {Arceneaux}, C. and {Areeda}, J.~S. and {Ashton}, G. and {Ast}, S. and {Aston}, S.~M. and {Aufmuth}, P. and {Aulbert}, C. and {Aylott}, B.~E. and {Babak}, S. and {Baker}, P.~T. and {Ballmer}, S.~W. and {Barayoga}, J.~C. and {Barbet}, M. and {Barclay}, S. and {Barish}, B.~C. and {Barker}, D. and {Barr}, B. and {Barsotti}, L. and {Bartlett}, J. and {Barton}, M.~A. and {Bartos}, I. and {Bassiri}, R. and {Batch}, J.~C. and {Baune}, C. and {Behnke}, B. and {Bell}, A.~S. and {Bell}, C. and {Benacquista}, M. and {Bergman}, J. and {Bergmann}, G. and {Berry}, C.~P.~L. and {Betzwieser}, J. and {Bhagwat}, S. and {Bhandare}, R. and {Bilenko}, I.~A. and {Billingsley}, G. and {Birch}, J. and {Biscans}, S. and {Biwer}, C. and {Blackburn}, J.~K. and {Blackburn}, L. and {Blair}, C.~D. and {Blair}, D. and {Bock}, O. and {Bodiya}, T.~P. and {Bojtos}, P. and {Bond}, C. and {Bork}, R. and {Born}, M. and {Bose}, Sukanta and {Brady}, P.~R. and {Braginsky}, V.~B. and {Brau}, J.~E. and {Bridges}, D.~O. and {Brinkmann}, M. and {Brooks}, A.~F. and {Brown}, D.~A. and {Brown}, D.~D. and {Brown}, N.~M. and {Buchman}, S. and {Buikema}, A. and {Buonanno}, A. and {Cadonati}, L. and {Calder{\'o}n Bustillo}, J. and {Camp}, J.~B. and {Cannon}, K.~C. and {Cao}, J. and {Capano}, C.~D. and {Caride}, S. and {Caudill}, S. and {Cavagli{\`a}}, M. and {Cepeda}, C. and {Chakraborty}, R. and {Chalermsongsak}, T. and {Chamberlin}, S.~J. and {Chao}, S. and {Charlton}, P. and {Chen}, Y. and {Cho}, H.~S. and {Cho}, M. and {Chow}, J.~H. and {Christensen}, N. and {Chu}, Q. and {Chung}, S. and {Ciani}, G. and {Clara}, F. and {Clark}, J.~A. and {Collette}, C. and {Cominsky}, L. and {Constancio}, Jr., M. and {Cook}, D. and {Corbitt}, T.~R. and {Cornish}, N. and {Corsi}, A. and {Costa}, C.~A. and {Coughlin}, M.~W. and {Countryman}, S. and {Couvares}, P. and {Coward}, D.~M. and {Cowart}, M.~J. and {Coyne}, D.~C. and {Coyne}, R. and {Craig}, K. and {Creighton}, J.~D.~E. and {Creighton}, T.~D. and {Cripe}, J. and {Crowder}, S.~G. and {Cumming}, A. and {Cunningham}, L. and {Cutler}, C. and {Dahl}, K. and {Dal Canton}, T. and {Damjanic}, M. and {Danilishin}, S.~L. and {Danzmann}, K. and {Dartez}, L. and {Dave}, I. and {Daveloza}, H. and {Davies}, G.~S. and {Daw}, E.~J. and {DeBra}, D. and {Del Pozzo}, W. and {Denker}, T. and {Dent}, T. and {Dergachev}, V. and {DeRosa}, R.~T. and {DeSalvo}, R. and {Dhurandhar}, S. and {D{\'\i}az}, M. and {Di Palma}, I. and {Dojcinoski}, G. and {Dominguez}, E. and {Donovan}, F. and {Dooley}, K.~L. and {Doravari}, S. and {Douglas}, R. and {Downes}, T.~P. and {Driggers}, J.~C. and {Du}, Z. and {Dwyer}, S. and {Eberle}, T. and {Edo}, T. and {Edwards}, M. and {Edwards}, M. and {Effler}, A. and {Eggenstein}, H.-B. and {Ehrens}, P. and {Eichholz}, J. and {Eikenberry}, S.~S. and {Essick}, R. and {Etzel}, T. and {Evans}, M. and {Evans}, T. and {Factourovich}, M. and {Fairhurst}, S. and {Fan}, X. and {Fang}, Q. and {Farr}, B. and {Farr}, W.~M. and {Favata}, M. and {Fays}, M. and {Fehrmann}, H. and {Fejer}, M.~M. and {Feldbaum}, D. and {Ferreira}, E.~C. and {Fisher}, R.~P. and {Frei}, Z. and {Freise}, A. and {Frey}, R. and {Fricke}, T.~T. and {Fritschel}, P. and {Frolov}, V.~V. and {Fuentes-Tapia}, S. and {Fulda}, P. and {Fyffe}, M. and {Gair}, J.~R.},
        title = "{Advanced LIGO}",
      journal = {Classical and Quantum Gravity},
         year = 2015,
        month = apr,
       volume = {32},
       number = {7},
          eid = {074001},
        pages = {074001},
          doi = {10.1088/0264-9381/32/7/074001},
archivePrefix = {arXiv},
       eprint = {1411.4547},
 primaryClass = {gr-qc},
       adsurl = {https://ui.adsabs.harvard.edu/abs/2015CQGra..32g4001L}
}

@article{AdvancedVirgo,
doi = {10.1088/0264-9381/32/2/024001},
url = {https://doi.org/10.1088/0264-9381/32/2/024001},
year = {2014},
month = {dec},
publisher = {IOP Publishing},
volume = {32},
number = {2},
pages = {024001},
author = {{The Virgo Collaboration} and Acernese, F and Agathos, M and Agatsuma, K and Aisa, D and Allemandou, N and Allocca, A and Amarni, J and Astone, P and Balestri, G and Ballardin, G and Barone, F and Baronick, J-P and Barsuglia, M and Basti, A and Basti, F and Bauer, Th S and Bavigadda, V and Bejger, M and Beker, M G and Belczynski, C and Bersanetti, D and Bertolini, A and Bitossi, M and Bizouard, M A and Bloemen, S and Blom, M and Boer, M and Bogaert, G and Bondi, D and Bondu, F and Bonelli, L and Bonnand, R and Boschi, V and Bosi, L and Bouedo, T and Bradaschia, C and Branchesi, M and Briant, T and Brillet, A and Brisson, V and Bulik, T and Bulten, H J and Buskulic, D and Buy, C and Cagnoli, G and Calloni, E and Campeggi, C and Canuel, B and Carbognani, F and Cavalier, F and Cavalieri, R and Cella, G and Cesarini, E and Mottin, E Chassande- and Chincarini, A and Chiummo, A and Chua, S and Cleva, F and Coccia, E and Cohadon, P-F and Colla, A and Colombini, M and Conte, A and Coulon, J-P and Cuoco, E and Dalmaz, A and D’Antonio, S and Dattilo, V and Davier, M and Day, R and Debreczeni, G and Degallaix, J and Deléglise, S and Pozzo, W Del and Dereli, H and Rosa, R De and Fiore, L Di and Lieto, A Di and Virgilio, A Di and Doets, M and Dolique, V and Drago, M and Ducrot, M and Endrőczi, G and Fafone, V and Farinon, S and Ferrante, I and Ferrini, F and Fidecaro, F and Fiori, I and Flaminio, R and Fournier, J-D and Franco, S and Frasca, S and Frasconi, F and Gammaitoni, L and Garufi, F and Gaspard, M and Gatto, A and Gemme, G and Gendre, B and Genin, E and Gennai, A and Ghosh, S and Giacobone, L and Giazotto, A and Gouaty, R and Granata, M and Greco, G and Groot, P and Guidi, G M and Harms, J and Heidmann, A and Heitmann, H and Hello, P and Hemming, G and Hennes, E and Hofman, D and Jaranowski, P and Jonker, R J G and Kasprzack, M and Kéfélian, F and Kowalska, I and Kraan, M and Królak, A and Kutynia, A and Lazzaro, C and Leonardi, M and Leroy, N and Letendre, N and Li, T G F and Lieunard, B and Lorenzini, M and Loriette, V and Losurdo, G and Magazzù, C and Majorana, E and Maksimovic, I and Malvezzi, V and Man, N and Mangano, V and Mantovani, M and Marchesoni, F and Marion, F and Marque, J and Martelli, F and Martellini, L and Masserot, A and Meacher, D and Meidam, J and Mezzani, F and Michel, C and Milano, L and Minenkov, Y and Moggi, A and Mohan, M and Montani, M and Morgado, N and Mours, B and Mul, F and Nagy, M F and Nardecchia, I and Naticchioni, L and Nelemans, G and Neri, I and Neri, M and Nocera, F and Pacaud, E and Palomba, C and Paoletti, F and Paoli, A and Pasqualetti, A and Passaquieti, R and Passuello, D and Perciballi, M and Petit, S and Pichot, M and Piergiovanni, F and Pillant, G and Piluso, A and Pinard, L and Poggiani, R and Prijatelj, M and Prodi, G A and Punturo, M and Puppo, P and Rabeling, D S and Rácz, I and Rapagnani, P and Razzano, M and Re, V and Regimbau, T and Ricci, F and Robinet, F and Rocchi, A and Rolland, L and Romano, R and Rosińska, D and Ruggi, P and Saracco, E and Sassolas, B and Schimmel, F and Sentenac, D and Sequino, V and Shah, S and Siellez, K and Straniero, N and Swinkels, B and Tacca, M and Tonelli, M and Travasso, F and Turconi, M and Vajente, G and van Bakel, N and van Beuzekom, M and van den Brand, J F J and Van Den Broeck, C and van der Sluys, M V and van Heijningen, J and Vasúth, M and Vedovato, G and Veitch, J and Verkindt, D and Vetrano, F and Viceré, A and Vinet, J-Y and Visser, G and Vocca, H and Ward, R and Was, M and Wei, L-W and Yvert, M and żny, A Zadro and Zendri, J-P},
title = {Advanced Virgo: a second-generation interferometric gravitational wave detector},
journal = {Classical and Quantum Gravity}
}

@ARTICLE{KAGRA,
       author = {{The KAGRA Collaboration} and {Akutsu}, T. and {Ando}, M. and {Arai}, K. and {Arai}, Y. and {Araki}, S. and {Araya}, A. and {Aritomi}, N. and {Aso}, Y. and {Bae}, S. and {Bae}, Y. and {Baiotti}, L. and {Bajpai}, R. and {Barton}, M.~A. and {Cannon}, K. and {Capocasa}, E. and {Chan}, M. and {Chen}, C. and {Chen}, K. and {Chen}, Y. and {Chu}, H. and {Chu}, Y.-K. and {Eguchi}, S. and {Enomoto}, Y. and {Flaminio}, R. and {Fujii}, Y. and {Fukunaga}, M. and {Fukushima}, M. and {Ge}, G. and {Hagiwara}, A. and {Haino}, S. and {Hasegawa}, K. and {Hayakawa}, H. and {Hayama}, K. and {Himemoto}, Y. and {Hiranuma}, Y. and {Hirata}, N. and {Hirose}, E. and {Hong}, Z. and {Hsieh}, B.~H. and {Huang}, C.-Z. and {Huang}, P. and {Huang}, Y. and {Ikenoue}, B. and {Imam}, S. and {Inayoshi}, K. and {Inoue}, Y. and {Ioka}, K. and {Itoh}, Y. and {Izumi}, K. and {Jung}, K. and {Jung}, P. and {Kajita}, T. and {Kamiizumi}, M. and {Kanda}, N. and {Kang}, G. and {Kawaguchi}, K. and {Kawai}, N. and {Kawasaki}, T. and {Kim}, C. and {Kim}, J.~C. and {Kim}, W.~S. and {Kim}, Y.-M. and {Kimura}, N. and {Kita}, N. and {Kitazawa}, H. and {Kojima}, Y. and {Kokeyama}, K. and {Komori}, K. and {Kong}, A.~K.~H. and {Kotake}, K. and {Kozakai}, C. and {Kozu}, R. and {Kumar}, R. and {Kume}, J. and {Kuo}, C. and {Kuo}, H.-S. and {Kuroyanagi}, S. and {Kusayanagi}, K. and {Kwak}, K. and {Lee}, H.~K. and {Lee}, H.~W. and {Lee}, R. and {Leonardi}, M. and {Lin}, L.~C.-C. and {Lin}, C.-Y. and {Lin}, F.-L. and {Liu}, G.~C. and {Luo}, L.-W. and {Marchio}, M. and {Michimura}, Y. and {Mio}, N. and {Miyakawa}, O. and {Miyamoto}, A. and {Miyazaki}, Y. and {Miyo}, K. and {Miyoki}, S. and {Morisaki}, S. and {Moriwaki}, Y. and {Nagano}, K. and {Nagano}, S. and {Nakamura}, K. and {Nakano}, H. and {Nakano}, M. and {Nakashima}, R. and {Narikawa}, T. and {Negishi}, R. and {Ni}, W.-T. and {Nishizawa}, A. and {Obuchi}, Y. and {Ogaki}, W. and {Oh}, J.~J. and {Oh}, S.~H. and {Ohashi}, M. and {Ohishi}, N. and {Ohkawa}, M. and {Okutomi}, K. and {Oohara}, K. and {Ooi}, C.~P. and {Oshino}, S. and {Pan}, K. and {Pang}, H. and {Park}, J. and {Arellano}, F.~E. Pe{\~n}a and {Pinto}, I. and {Sago}, N. and {Saito}, S. and {Saito}, Y. and {Sakai}, K. and {Sakai}, Y. and {Sakuno}, Y. and {Sato}, S. and {Sato}, T. and {Sawada}, T. and {Sekiguchi}, T. and {Sekiguchi}, Y. and {Shibagaki}, S. and {Shimizu}, R. and {Shimoda}, T. and {Shimode}, K. and {Shinkai}, H. and {Shishido}, T. and {Shoda}, A. and {Somiya}, K. and {Son}, E.~J. and {Sotani}, H. and {Sugimoto}, R. and {Suzuki}, T. and {Suzuki}, T. and {Tagoshi}, H. and {Takahashi}, H. and {Takahashi}, R. and {Takamori}, A. and {Takano}, S. and {Takeda}, H. and {Takeda}, M. and {Tanaka}, H. and {Tanaka}, K. and {Tanaka}, K. and {Tanaka}, T. and {Tanaka}, T. and {Tanioka}, S. and {Tapia San Martin}, E.~N. and {Telada}, S. and {Tomaru}, T. and {Tomigami}, Y. and {Tomura}, T. and {Travasso}, F. and {Trozzo}, L. and {Tsang}, T. and {Tsubono}, K. and {Tsuchida}, S. and {Tsuzuki}, T. and {Tuyenbayev}, D. and {Uchikata}, N. and {Uchiyama}, T. and {Ueda}, A. and {Uehara}, T. and {Ueno}, K. and {Ueshima}, G. and {Uraguchi}, F. and {Ushiba}, T. and {van Putten}, M.~H.~P.~M. and {Vocca}, H. and {Wang}, J. and {Wu}, C. and {Wu}, H. and {Wu}, S. and {Xu}, W.-R. and {Yamada}, T. and {Yamamoto}, K. and {Yamamoto}, K. and {Yamamoto}, T. and {Yokogawa}, K. and {Yokoyama}, J. and {Yokozawa}, T. and {Yoshioka}, T. and {Yuzurihara}, H. and {Zeidler}, S. and {Zhao}, Y. and {Zhu}, Z.-H.},
        title = "{Overview of KAGRA: Detector design and construction history}",
      journal = {Progress of Theoretical and Experimental Physics},
         year = 2021,
        month = may,
       volume = {2021},
       number = {5},
          eid = {05A101},
        pages = {05A101},
          doi = {10.1093/ptep/ptaa125},
archivePrefix = {arXiv},
       eprint = {2005.05574},
 primaryClass = {physics.ins-det},
       adsurl = {https://ui.adsabs.harvard.edu/abs/2021PTEP.2021eA101A}
}

@ARTICLE{GWCosmo1,
       author = {{The LIGO Scientific Collaboration} and {the Virgo Collaboration} and {the KAGRA Collaboration} and {Abac}, A.~G. and {Abe}, A. and {Abouelfettouh}, I. and {Acernese}, F. and {Ackley}, K. and {Adam}, A. and {Adhicary}, S. and {Adhikari}, D. and {Adhikari}, R.~X. and {Adkins}, V.~K. and {Afroz}, S. and {Agapito}, A. and {Agarwal}, D. and {Agathos}, M. and {Aggarwal}, N. and {Aggarwal}, S. and {Aguiar}, O.~D. and {Ahrend}, I.-L. and {Aiello}, L. and {Ain}, A. and {Ajith}, P. and {Akutsu}, T. and {Albers}, L. and {Ali}, W. and {Al-Kershi}, S. and {Allene}, C. and {Allocca}, A. and {Al-Shammari}, S. and {Alvarez}, J.~A. and {Alvarez-Lopez}, S. and {Amar}, W. and {Amarasinghe}, O. and {Amato}, A. and {Amicucci}, F. and {Amra}, C. and {Anand}, A.~B. and {Anand}, C. and {Ananyeva}, A. and {Anderson}, S.~B. and {Anderson}, W.~G. and {Andia}, M. and {Ando}, M. and {Andrade-Oliveira}, F. and {Andr{\'e}s-Carcasona}, M. and {Andrey}, J.~L. and {Andri{\'c}}, T. and {Anglin}, J. and {Anna}, J. and {Antelis}, J.~M. and {Antier}, S. and {Aoki}, T. and {Aoumi}, M. and {Appavuravther}, E.~Z. and {Appelt}, E.~A. and {Appert}, S. and {Apple}, S.~K. and {Arai}, K. and {Araya}, A. and {Araya}, M.~C. and {Arca Sedda}, M. and {Arciprete}, F. and {Areeda}, J.~S. and {Aritomi}, N. and {Armato}, F. and {Armstrong}, S. and {Arnaud}, N. and {Arogeti}, M. and {Aronson}, S.~M. and {Ashton}, G. and {Aso}, Y. and {Asprea}, L. and {Assiduo}, M. and {Assis de Souza Melo}, S. and {Aston}, S.~M. and {Astone}, P. and {Aswathi}, P.~S. and {Attadio}, F. and {Aubin}, F. and {AultONeal}, K. and {Avallone}, G. and {Avdeev}, N. and {Avila}, E.~A. and {Babak}, S. and {Badger}, C. and {Bae}, S. and {Bagnasco}, S. and {Baimukhametova}, S. and {Baiotti}, L. and {Baka}, T. and {Baker}, K.~A. and {Baker}, T. and {Balbi}, G. and {Baldi}, G. and {Baldicchi}, N. and {Ball}, M. and {Ballardin}, G. and {Ballelli}, M. and {Ballmer}, S.~W. and {Banagiri}, S. and {Banerjee}, B. and {Bankar}, D. and {Baptiste}, T.~M. and {Baral}, P. and {Baratti}, M. and {Barayoga}, J.~C. and {Baric}, K. and {Barish}, B.~C. and {Barker}, D. and {Barman}, N. and {Barone}, F. and {Barr}, B. and {Barrios}, M. and {Barsotti}, L. and {Barsuglia}, M. and {Barta}, D. and {Barton}, M.~A. and {Bartos}, I. and {Basalaev}, A. and {Bassiri}, R. and {Basti}, A. and {Bawaj}, M. and {Bayley}, J.~C. and {Baylor}, A.~C. and {Baynard}, II, P.~A. and {Bazzan}, M. and {Bedakihale}, V.~M. and {Beirnaert}, F. and {Bejger}, M. and {Bell}, A.~S. and {Bellani}, C. and {Bellie}, D.~S. and {Beltran-Martinez}, D. and {Benedetti}, E. and {Benoit}, W. and {Bentara}, I. and {Ben Yaala}, M. and {Bera}, S. and {Bergamin}, F. and {Berger}, B.~K. and {Beroiz}, M. and {Berry}, C.~P.~L. and {Berry}, I. and {Bersanetti}, D. and {Bertheas}, T. and {Bertolini}, A. and {Betzwieser}, J. and {Beveridge}, D. and {Bevins}, N. and {Bezerra-Sobrinho}, J. and {Bhandare}, R. and {Bhatt}, R. and {Bhattacharjee}, A. and {Bhattacharjee}, D. and {Bhattacharyya}, S. and {Bhaumik}, S. and {Biancalana}, V. and {Bianchi}, F. and {Bilenko}, I.~A. and {Bilicki}, M. and {Billingsley}, G. and {Binetti}, A. and {Bini}, S. and {Biot}, S. and {Birnholtz}, O. and {Biscoveanu}, S. and {Bisht}, A. and {Bitossi}, M. and {Bizouard}, M.-A. and {Blaber}, S. and {Blackburn}, J.~K. and {Blagg}, L.~A. and {Blair}, C.~D. and {Blair}, D.~G. and {Bloch}, M. and {Bode}, N. and {Boettner}, N. and {Bogdan}, P. and {Boileau}, G. and {Boldrini}, M. and {Bolingbroke}, G.~N. and {Bonavena}, L.~D. and {Bonhomme}, V.~A. and {Bonilla}, E. and {Bonilla}, M.~S. and {Bonino}, A. and {Bonnand}, R. and {Borchers}, A. and {Borghi}, N. and {Boschi}, V. and {Bose}, S. and {Bossilkov}, V. and {Bothra}, Y. and {Boudon}, A. and {Boybeyi}, T.~D. and {Boyle}, M. and {Bozzi}, A. and {Bradaschia}, C.},
        title = "{GWTC-5.0: Constraints on the Cosmic Expansion Rate and Modified Gravitational-wave Propagation}",
      journal = {arXiv e-prints},
         year = 2026,
        month = may,
          eid = {arXiv:2605.27227},
        pages = {arXiv:2605.27227},
          doi = {10.48550/arXiv.2605.27227},
archivePrefix = {arXiv},
       eprint = {2605.27227},
 primaryClass = {astro-ph.CO},
       adsurl = {https://ui.adsabs.harvard.edu/abs/2026arXiv260527227T}
}

@ARTICLE{TGR1,
       author = {{The LIGO Scientific Collaboration} and {the Virgo Collaboration} and {the KAGRA Collaboration} and {Abac}, A.~G. and {Abouelfettouh}, I. and {Acernese}, F. and {Ackley}, K. and {Adamcewicz}, C. and {Adhicary}, S. and {Adhikari}, D. and {Adhikari}, N. and {Adhikari}, R.~X. and {Adkins}, V.~K. and {Afroz}, S. and {Agapito}, A. and {Agarwal}, D. and {Agathos}, M. and {Aggarwal}, N. and {Aggarwal}, S. and {Aguiar}, O.~D. and {Ahrend}, I.-L. and {Aiello}, L. and {Ain}, A. and {Ajith}, P. and {Akutsu}, T. and {Albanesi}, S. and {Ali}, W. and {Al-Kershi}, S. and {All{\'e}n{\'e}}, C. and {Allocca}, A. and {Al-Shammari}, S. and {Altin}, P.~A. and {Alvarez-Lopez}, S. and {Amar}, W. and {Amarasinghe}, O. and {Amato}, A. and {Amicucci}, F. and {Amra}, C. and {Ananyeva}, A. and {Anderson}, S.~B. and {Anderson}, W.~G. and {Andia}, M. and {Ando}, M. and {Andr{\'e}s-Carcasona}, M. and {Andri{\'c}}, T. and {Anglin}, J. and {Ansoldi}, S. and {Antelis}, J.~M. and {Antier}, S. and {Aoumi}, M. and {Appavuravther}, E.~Z. and {Appert}, S. and {Apple}, S.~K. and {Arai}, K. and {Araya}, A. and {Araya}, M.~C. and {Arca Sedda}, M. and {Areeda}, J.~S. and {Aritomi}, N. and {Armato}, F. and {Armstrong}, S. and {Arnaud}, N. and {Arogeti}, M. and {Aronson}, S.~M. and {Arun}, K.~G. and {Ashton}, G. and {Aso}, Y. and {Asprea}, L. and {Assiduo}, M. and {Assis de Souza Melo}, S. and {Aston}, S.~M. and {Astone}, P. and {Attadio}, F. and {Aubin}, F. and {AultONeal}, K. and {Avallone}, G. and {Avila}, E.~A. and {Babak}, S. and {Badger}, C. and {Bae}, S. and {Bagnasco}, S. and {Baiotti}, L. and {Bajpai}, R. and {Baka}, T. and {Baker}, A.~M. and {Baker}, K.~A. and {Baker}, T. and {Baldi}, G. and {Baldicchi}, N. and {Ball}, M. and {Ballardin}, G. and {Ballmer}, S.~W. and {Banagiri}, S. and {Banerjee}, B. and {Bankar}, D. and {Baptiste}, T.~M. and {Baral}, P. and {Baratti}, M. and {Barayoga}, J.~C. and {Barish}, B.~C. and {Barker}, D. and {Barman}, N. and {Barneo}, P. and {Barone}, F. and {Barr}, B. and {Barsotti}, L. and {Barsuglia}, M. and {Barta}, D. and {Bartoletti}, A.~M. and {Barton}, M.~A. and {Bartos}, I. and {Basalaev}, A. and {Bassiri}, R. and {Basti}, A. and {Bawaj}, M. and {Baxi}, P. and {Bayley}, J.~C. and {Baylor}, A.~C. and {Baynard}, II, P.~A. and {Bazzan}, M. and {Bedakihale}, V.~M. and {Beirnaert}, F. and {Bejger}, M. and {Belardinelli}, D. and {Bell}, A.~S. and {Bellie}, D.~S. and {Bellizzi}, L. and {Benoit}, W. and {Bentara}, I. and {Bentley}, J.~D. and {Ben Yaala}, M. and {Bera}, S. and {Bergamin}, F. and {Berger}, B.~K. and {Bernuzzi}, S. and {Beroiz}, M. and {Berry}, C.~P.~L. and {Bersanetti}, D. and {Bertheas}, T. and {Bertolini}, A. and {Betzwieser}, J. and {Beveridge}, D. and {Bevilacqua}, G. and {Bevins}, N. and {Bhagwat}, S. and {Bhandare}, R. and {Bhat}, S.~A. and {Bhatt}, R. and {Bhattacharjee}, D. and {Bhattacharyya}, S. and {Bhaumik}, S. and {Biancalana}, V. and {Bianchi}, A. and {Bilenko}, I.~A. and {Billingsley}, G. and {Binetti}, A. and {Bini}, S. and {Binu}, C. and {Biot}, S. and {Birnholtz}, O. and {Biscoveanu}, S. and {Bisht}, A. and {Bitossi}, M. and {Bizouard}, M.-A. and {Blaber}, S. and {Blackburn}, J.~K. and {Blagg}, L.~A. and {Blair}, C.~D. and {Blair}, D.~G. and {Bode}, N. and {Boettner}, N. and {Boileau}, G. and {Boldrini}, M. and {Bolingbroke}, G.~N. and {Bolliand}, A. and {Bonavena}, L.~D. and {Bondarescu}, R. and {Bondu}, F. and {Bonilla}, E. and {Bonilla}, M.~S. and {Bonino}, A. and {Bonnand}, R. and {Borchers}, A. and {Boschi}, V. and {Bose}, S. and {Bossilkov}, V. and {Bothra}, Y. and {Boudon}, A. and {Bourg}, L. and {Boyle}, M. and {Bozzi}, A. and {Bradaschia}, C. and {Brady}, P.~R. and {Branch}, A. and {Branchesi}, M. and {Braun}, I. and {Briant}, T. and {Brillet}, A. and {Brinkmann}, M. and {Brockill}, P.},
        title = "{GWTC-4.0: Tests of General Relativity. I. Overview and General Tests}",
      journal = {arXiv e-prints},
         year = 2026,
        month = mar,
          eid = {arXiv:2603.19019},
        pages = {arXiv:2603.19019},
          doi = {10.48550/arXiv.2603.19019},
archivePrefix = {arXiv},
       eprint = {2603.19019},
 primaryClass = {gr-qc},
       adsurl = {https://ui.adsabs.harvard.edu/abs/2026arXiv260319019T}
}

@ARTICLE{TGR2,
       author = {{The LIGO Scientific Collaboration} and {the Virgo Collaboration} and {the KAGRA Collaboration} and {Abac}, A.~G. and {Abouelfettouh}, I. and {Acernese}, F. and {Ackley}, K. and {Adamcewicz}, C. and {Adhicary}, S. and {Adhikari}, D. and {Adhikari}, N. and {Adhikari}, R.~X. and {Adkins}, V.~K. and {Afroz}, S. and {Agapito}, A. and {Agarwal}, D. and {Agathos}, M. and {Aggarwal}, N. and {Aggarwal}, S. and {Aguiar}, O.~D. and {Ahrend}, I.-L. and {Aiello}, L. and {Ain}, A. and {Ajith}, P. and {Akutsu}, T. and {Albanesi}, S. and {Ali}, W. and {Al-Kershi}, S. and {All{\'e}n{\'e}}, C. and {Allocca}, A. and {Al-Shammari}, S. and {Altin}, P.~A. and {Alvarez-Lopez}, S. and {Amar}, W. and {Amarasinghe}, O. and {Amato}, A. and {Amicucci}, F. and {Amra}, C. and {Ananyeva}, A. and {Anderson}, S.~B. and {Anderson}, W.~G. and {Andia}, M. and {Ando}, M. and {Andr{\'e}s-Carcasona}, M. and {Andri{\'c}}, T. and {Anglin}, J. and {Ansoldi}, S. and {Antelis}, J.~M. and {Antier}, S. and {Aoumi}, M. and {Appavuravther}, E.~Z. and {Appert}, S. and {Apple}, S.~K. and {Arai}, K. and {Araya}, A. and {Araya}, M.~C. and {Arca Sedda}, M. and {Areeda}, J.~S. and {Aritomi}, N. and {Armato}, F. and {Armstrong}, S. and {Arnaud}, N. and {Arogeti}, M. and {Aronson}, S.~M. and {Arun}, K.~G. and {Ashton}, G. and {Aso}, Y. and {Asprea}, L. and {Assiduo}, M. and {Assis de Souza Melo}, S. and {Aston}, S.~M. and {Astone}, P. and {Attadio}, F. and {Aubin}, F. and {AultONeal}, K. and {Avallone}, G. and {Avila}, E.~A. and {Babak}, S. and {Badger}, C. and {Bae}, S. and {Bagnasco}, S. and {Baiotti}, L. and {Bajpai}, R. and {Baka}, T. and {Baker}, A.~M. and {Baker}, K.~A. and {Baker}, T. and {Baldi}, G. and {Baldicchi}, N. and {Ball}, M. and {Ballardin}, G. and {Ballmer}, S.~W. and {Banagiri}, S. and {Banerjee}, B. and {Bankar}, D. and {Baptiste}, T.~M. and {Baral}, P. and {Baratti}, M. and {Barayoga}, J.~C. and {Barish}, B.~C. and {Barker}, D. and {Barman}, N. and {Barneo}, P. and {Barone}, F. and {Barr}, B. and {Barsotti}, L. and {Barsuglia}, M. and {Barta}, D. and {Bartoletti}, A.~M. and {Barton}, M.~A. and {Bartos}, I. and {Basalaev}, A. and {Bassiri}, R. and {Basti}, A. and {Bawaj}, M. and {Baxi}, P. and {Bayley}, J.~C. and {Baylor}, A.~C. and {Baynard}, II, P.~A. and {Bazzan}, M. and {Bedakihale}, V.~M. and {Beirnaert}, F. and {Bejger}, M. and {Belardinelli}, D. and {Bell}, A.~S. and {Bellie}, D.~S. and {Bellizzi}, L. and {Benoit}, W. and {Bentara}, I. and {Bentley}, J.~D. and {Ben Yaala}, M. and {Bera}, S. and {Bergamin}, F. and {Berger}, B.~K. and {Bernuzzi}, S. and {Beroiz}, M. and {Berry}, C.~P.~L. and {Bersanetti}, D. and {Bertheas}, T. and {Bertolini}, A. and {Betzwieser}, J. and {Beveridge}, D. and {Bevilacqua}, G. and {Bevins}, N. and {Bhagwat}, S. and {Bhandare}, R. and {Bhat}, S.~A. and {Bhatt}, R. and {Bhattacharjee}, D. and {Bhattacharyya}, S. and {Bhaumik}, S. and {Biancalana}, V. and {Bianchi}, A. and {Bilenko}, I.~A. and {Billingsley}, G. and {Binetti}, A. and {Bini}, S. and {Binu}, C. and {Biot}, S. and {Birnholtz}, O. and {Biscoveanu}, S. and {Bisht}, A. and {Bitossi}, M. and {Bizouard}, M.-A. and {Blaber}, S. and {Blackburn}, J.~K. and {Blagg}, L.~A. and {Blair}, C.~D. and {Blair}, D.~G. and {Bode}, N. and {Boettner}, N. and {Boileau}, G. and {Boldrini}, M. and {Bolingbroke}, G.~N. and {Bolliand}, A. and {Bonavena}, L.~D. and {Bondarescu}, R. and {Bondu}, F. and {Bonilla}, E. and {Bonilla}, M.~S. and {Bonino}, A. and {Bonnand}, R. and {Borchers}, A. and {Boschi}, V. and {Bose}, S. and {Bossilkov}, V. and {Bothra}, Y. and {Boudon}, A. and {Bourg}, L. and {Boyle}, M. and {Bozzi}, A. and {Bradaschia}, C. and {Brady}, P.~R. and {Branch}, A. and {Branchesi}, M. and {Braun}, I. and {Briant}, T. and {Brillet}, A. and {Brinkmann}, M. and {Brockill}, P.},
        title = "{GWTC-4.0: Tests of General Relativity. II. Parameterized Tests}",
      journal = {arXiv e-prints},
         year = 2026,
        month = mar,
          eid = {arXiv:2603.19020},
        pages = {arXiv:2603.19020},
          doi = {10.48550/arXiv.2603.19020},
archivePrefix = {arXiv},
       eprint = {2603.19020},
 primaryClass = {gr-qc},
       adsurl = {https://ui.adsabs.harvard.edu/abs/2026arXiv260319020T}
}

@ARTICLE{TGR3,
       author = {{The LIGO Scientific Collaboration} and {the Virgo Collaboration} and {the KAGRA Collaboration} and {Abac}, A.~G. and {Abouelfettouh}, I. and {Acernese}, F. and {Ackley}, K. and {Adamcewicz}, C. and {Adhicary}, S. and {Adhikari}, D. and {Adhikari}, N. and {Adhikari}, R.~X. and {Adkins}, V.~K. and {Afroz}, S. and {Agapito}, A. and {Agarwal}, D. and {Agathos}, M. and {Aggarwal}, N. and {Aggarwal}, S. and {Aguiar}, O.~D. and {Ahrend}, I.-L. and {Aiello}, L. and {Ain}, A. and {Ajith}, P. and {Akutsu}, T. and {Albanesi}, S. and {Ali}, W. and {Al-Kershi}, S. and {All{\'e}n{\'e}}, C. and {Allocca}, A. and {Al-Shammari}, S. and {Altin}, P.~A. and {Alvarez-Lopez}, S. and {Amar}, W. and {Amarasinghe}, O. and {Amato}, A. and {Amicucci}, F. and {Amra}, C. and {Ananyeva}, A. and {Anderson}, S.~B. and {Anderson}, W.~G. and {Andia}, M. and {Ando}, M. and {Andr{\'e}s-Carcasona}, M. and {Andri{\'c}}, T. and {Anglin}, J. and {Ansoldi}, S. and {Antelis}, J.~M. and {Antier}, S. and {Aoumi}, M. and {Appavuravther}, E.~Z. and {Appert}, S. and {Apple}, S.~K. and {Arai}, K. and {Araya}, A. and {Araya}, M.~C. and {Arca Sedda}, M. and {Areeda}, J.~S. and {Aritomi}, N. and {Armato}, F. and {Armstrong}, S. and {Arnaud}, N. and {Arogeti}, M. and {Aronson}, S.~M. and {Ashton}, G. and {Aso}, Y. and {Asprea}, L. and {Assiduo}, M. and {Assis de Souza Melo}, S. and {Aston}, S.~M. and {Astone}, P. and {Attadio}, F. and {Aubin}, F. and {AultONeal}, K. and {Avallone}, G. and {Avila}, E.~A. and {Babak}, S. and {Badger}, C. and {Bae}, S. and {Bagnasco}, S. and {Baiotti}, L. and {Bajpai}, R. and {Baka}, T. and {Baker}, A.~M. and {Baker}, K.~A. and {Baker}, T. and {Baldi}, G. and {Baldicchi}, N. and {Ball}, M. and {Ballardin}, G. and {Ballmer}, S.~W. and {Banagiri}, S. and {Banerjee}, B. and {Bankar}, D. and {Baptiste}, T.~M. and {Baral}, P. and {Baratti}, M. and {Barayoga}, J.~C. and {Barish}, B.~C. and {Barker}, D. and {Barman}, N. and {Barneo}, P. and {Barone}, F. and {Barr}, B. and {Barsotti}, L. and {Barsuglia}, M. and {Barta}, D. and {Bartoletti}, A.~M. and {Barton}, M.~A. and {Bartos}, I. and {Basalaev}, A. and {Bassiri}, R. and {Basti}, A. and {Bawaj}, M. and {Baxi}, P. and {Bayley}, J.~C. and {Baylor}, A.~C. and {Baynard}, II, P.~A. and {Bazzan}, M. and {Bedakihale}, V.~M. and {Beirnaert}, F. and {Bejger}, M. and {Belardinelli}, D. and {Bell}, A.~S. and {Bellie}, D.~S. and {Bellizzi}, L. and {Benoit}, W. and {Bentara}, I. and {Bentley}, J.~D. and {Ben Yaala}, M. and {Bera}, S. and {Bergamin}, F. and {Berger}, B.~K. and {Bernuzzi}, S. and {Beroiz}, M. and {Berry}, C.~P.~L. and {Bersanetti}, D. and {Bertheas}, T. and {Bertolini}, A. and {Betzwieser}, J. and {Beveridge}, D. and {Bevilacqua}, G. and {Bevins}, N. and {Bhagwat}, S. and {Bhandare}, R. and {Bhatt}, R. and {Bhattacharjee}, D. and {Bhattacharyya}, S. and {Bhaumik}, S. and {Biancalana}, V. and {Bianchi}, A. and {Bilenko}, I.~A. and {Billingsley}, G. and {Binetti}, A. and {Bini}, S. and {Binu}, C. and {Biot}, S. and {Birnholtz}, O. and {Biscoveanu}, S. and {Bisht}, A. and {Bitossi}, M. and {Bizouard}, M.-A. and {Blaber}, S. and {Blackburn}, J.~K. and {Blagg}, L.~A. and {Blair}, C.~D. and {Blair}, D.~G. and {Bode}, N. and {Boettner}, N. and {Boileau}, G. and {Boldrini}, M. and {Bolingbroke}, G.~N. and {Bolliand}, A. and {Bonavena}, L.~D. and {Bondarescu}, R. and {Bondu}, F. and {Bonilla}, E. and {Bonilla}, M.~S. and {Bonino}, A. and {Bonnand}, R. and {Borchers}, A. and {Boschi}, V. and {Bose}, S. and {Bossilkov}, V. and {Bothra}, Y. and {Boudon}, A. and {Bourg}, L. and {Boyle}, M. and {Bozzi}, A. and {Bradaschia}, C. and {Brady}, P.~R. and {Branch}, A. and {Branchesi}, M. and {Braun}, I. and {Briant}, T. and {Brillet}, A. and {Brinkmann}, M. and {Brockill}, P. and {Brockmueller}, E. and {Brooks}, A.~F.},
        title = "{GWTC-4.0: Tests of General Relativity. III. Tests of the Remnants}",
      journal = {arXiv e-prints},
         year = 2026,
        month = mar,
          eid = {arXiv:2603.19021},
        pages = {arXiv:2603.19021},
          doi = {10.48550/arXiv.2603.19021},
archivePrefix = {arXiv},
       eprint = {2603.19021},
 primaryClass = {gr-qc},
       adsurl = {https://ui.adsabs.harvard.edu/abs/2026arXiv260319021T}
}

@ARTICLE{GWpop1,
       author = {{The LIGO Scientific Collaboration} and {the Virgo Collaboration} and {the KAGRA Collaboration} and {Abac}, A.~G. and {Abe}, A. and {Abouelfettouh}, I. and {Acernese}, F. and {Ackley}, K. and {Adam}, A. and {Adhicary}, S. and {Adhikari}, D. and {Adhikari}, R.~X. and {Adkins}, V.~K. and {Afroz}, S. and {Agapito}, A. and {Agarwal}, D. and {Agathos}, M. and {Aggarwal}, N. and {Aggarwal}, S. and {Aguiar}, O.~D. and {Ahrend}, I.-L. and {Aiello}, L. and {Ain}, A. and {Ajith}, P. and {Akutsu}, T. and {Albers}, L. and {Ali}, W. and {Al-Kershi}, S. and {Allene}, C. and {Allocca}, A. and {Al-Shammari}, S. and {Alvarez}, J.~A. and {Alvarez-Lopez}, S. and {Amar}, W. and {Amarasinghe}, O. and {Amato}, A. and {Amicucci}, F. and {Amra}, C. and {Anand}, A.~B. and {Anand}, C. and {Ananyeva}, A. and {Anderson}, S.~B. and {Anderson}, W.~G. and {Andia}, M. and {Ando}, M. and {Andrade-Oliveira}, F. and {Andr{\'e}s-Carcasona}, M. and {Andrey}, J.~L. and {Andri{\'c}}, T. and {Anglin}, J. and {Anna}, J. and {Antelis}, J.~M. and {Antier}, S. and {Antonini}, F. and {Aoki}, T. and {Aoumi}, M. and {Appavuravther}, E.~Z. and {Appelt}, E.~A. and {Appert}, S. and {Apple}, S.~K. and {Arai}, K. and {Araya}, A. and {Araya}, M.~C. and {Arca Sedda}, M. and {Arciprete}, F. and {Areeda}, J.~S. and {Aritomi}, N. and {Armato}, F. and {Armstrong}, S. and {Arnaud}, N. and {Arogeti}, M. and {Aronson}, S.~M. and {Ashton}, G. and {Aso}, Y. and {Asprea}, L. and {Assiduo}, M. and {Assis de Souza Melo}, S. and {Aston}, S.~M. and {Astone}, P. and {Aswathi}, P.~S. and {Attadio}, F. and {Aubin}, F. and {AultONeal}, K. and {Avallone}, G. and {Avdeev}, N. and {Avila}, E.~A. and {Babak}, S. and {Badger}, C. and {Bae}, S. and {Bagnasco}, S. and {Baimukhametova}, S. and {Baiotti}, L. and {Baka}, T. and {Baker}, K.~A. and {Baker}, T. and {Balbi}, G. and {Baldi}, G. and {Baldicchi}, N. and {Ball}, M. and {Ballardin}, G. and {Ballelli}, M. and {Ballmer}, S.~W. and {Banagiri}, S. and {Banerjee}, B. and {Bankar}, D. and {Baptiste}, T.~M. and {Baral}, P. and {Baratti}, M. and {Barayoga}, J.~C. and {Baric}, K. and {Barish}, B.~C. and {Barker}, D. and {Barman}, N. and {Barone}, F. and {Barr}, B. and {Barrios}, M. and {Barsotti}, L. and {Barsuglia}, M. and {Barta}, D. and {Barton}, M.~A. and {Bartos}, I. and {Basalaev}, A. and {Bassiri}, R. and {Basti}, A. and {Bawaj}, M. and {Bayley}, J.~C. and {Baylor}, A.~C. and {Baynard}, II, P.~A. and {Bazzan}, M. and {Bedakihale}, V.~M. and {Beirnaert}, F. and {Bejger}, M. and {Bell}, A.~S. and {Bellani}, C. and {Bellie}, D.~S. and {Beltran-Martinez}, D. and {Benedetti}, E. and {Benoit}, W. and {Bentara}, I. and {Ben Yaala}, M. and {Bera}, S. and {Bergamin}, F. and {Berger}, B.~K. and {Beroiz}, M. and {Berry}, C.~P.~L. and {Berry}, I. and {Bersanetti}, D. and {Bertheas}, T. and {Bertolini}, A. and {Betzwieser}, J. and {Beveridge}, D. and {Bevins}, N. and {Bezerra-Sobrinho}, J. and {Bhandare}, R. and {Bhatt}, R. and {Bhattacharjee}, A. and {Bhattacharjee}, D. and {Bhattacharyya}, S. and {Bhaumik}, S. and {Biancalana}, V. and {Bianchi}, F. and {Bilenko}, I.~A. and {Bilicki}, M. and {Billingsley}, G. and {Binetti}, A. and {Bini}, S. and {Biot}, S. and {Birnholtz}, O. and {Biscoveanu}, S. and {Bisht}, A. and {Bitossi}, M. and {Bizouard}, M.-A. and {Blaber}, S. and {Blackburn}, J.~K. and {Blagg}, L.~A. and {Blair}, C.~D. and {Blair}, D.~G. and {Bloch}, M. and {Bode}, N. and {Boettner}, N. and {Bogdan}, P. and {Boileau}, G. and {Boldrini}, M. and {Bolingbroke}, G.~N. and {Bonavena}, L.~D. and {Bonhomme}, V.~A. and {Bonilla}, E. and {Bonilla}, M.~S. and {Bonino}, A. and {Bonnand}, R. and {Borchers}, A. and {Borghi}, N. and {Boschi}, V. and {Bose}, S. and {Bossilkov}, V. and {Bothra}, Y. and {Boudon}, A. and {Boybeyi}, T.~D. and {Boyle}, M. and {Bozzi}, A.},
        title = "{GWTC-5.0: Population Properties of Merging Compact Binaries}",
      journal = {arXiv e-prints},
         year = 2026,
        month = may,
          eid = {arXiv:2605.27226},
        pages = {arXiv:2605.27226},
          doi = {10.48550/arXiv.2605.27226},
archivePrefix = {arXiv},
       eprint = {2605.27226},
 primaryClass = {astro-ph.HE},
       adsurl = {https://ui.adsabs.harvard.edu/abs/2026arXiv260527226T}
}

@ARTICLE{CW1,
       author = {{Sieniawska}, Magdalena and {Bejger}, Micha{\l}},
        title = "{Continuous Gravitational Waves from Neutron Stars: Current Status and Prospects}",
      journal = {Universe},
         year = 2019,
        month = oct,
       volume = {5},
       number = {11},
          eid = {217},
        pages = {217},
          doi = {10.3390/universe5110217},
archivePrefix = {arXiv},
       eprint = {1909.12600},
 primaryClass = {astro-ph.HE},
       adsurl = {https://ui.adsabs.harvard.edu/abs/2019Univ....5..217S}
}

@INPROCEEDINGS{CW2,
       author = {{Glampedakis}, Kostas and {Gualtieri}, Leonardo},
        title = "{Gravitational Waves from Single Neutron Stars: An Advanced Detector Era Survey}",
    booktitle = {Astrophysics and Space Science Library},
         year = 2018,
       editor = {{Rezzolla}, Luciano and {Pizzochero}, Pierre and {Jones}, David Ian and {Rea}, Nanda and {Vida{\~n}a}, Isaac},
       series = {Astrophysics and Space Science Library},
       volume = {457},
        month = jan,
        pages = {673},
          doi = {10.1007/978-3-319-97616-7_12},
archivePrefix = {arXiv},
       eprint = {1709.07049},
 primaryClass = {astro-ph.HE},
       adsurl = {https://ui.adsabs.harvard.edu/abs/2018ASSL..457..673G}
}

@ARTICLE{DS1,
       author = {{Piccinni}, Ornella J. and {Astone}, P. and {D'Antonio}, S. and {Frasca}, S. and {Intini}, G. and {La Rosa}, I. and {Leaci}, P. and {Mastrogiovanni}, S. and {Miller}, A. and {Palomba}, C.},
        title = "{Directed search for continuous gravitational-wave signals from the Galactic Center in the Advanced LIGO second observing run}",
      journal = {Physical Review D},
         year = 2020,
        month = apr,
       volume = {101},
       number = {8},
          eid = {082004},
        pages = {082004},
          doi = {10.1103/PhysRevD.101.082004},
archivePrefix = {arXiv},
       eprint = {1910.05097},
 primaryClass = {gr-qc},
       adsurl = {https://ui.adsabs.harvard.edu/abs/2020PhRvD.101h2004P}
}

@ARTICLE{DS2,
       author = {{The LIGO Scientific Collaboration} and {the Virgo Collaboration} and {the KAGRA Collaboration} and {Abbott}, R. and {Abe}, H. and {Acernese}, F. and {Ackley}, K. and {Adhikari}, N. and {Adhikari}, R.~X. and {Adkins}, V.~K. and {Adya}, V.~B. and {Affeldt}, C. and {Agarwal}, D. and {Agathos}, M. and {Agatsuma}, K. and {Aggarwal}, N. and {Aguiar}, O.~D. and {Aiello}, L. and {Ain}, A. and {Ajith}, P. and {Akutsu}, T. and {Albanesi}, S. and {Alfaidi}, R.~A. and {Allocca}, A. and {Altin}, P.~A. and {Amato}, A. and {Anand}, C. and {Anand}, S. and {Ananyeva}, A. and {Anderson}, S.~B. and {Anderson}, W.~G. and {Ando}, M. and {Andrade}, T. and {Andres}, N. and {Andr{\'e}s-Carcasona}, M. and {Andri{\'c}}, T. and {Angelova}, S.~V. and {Ansoldi}, S. and {Antelis}, J.~M. and {Antier}, S. and {Apostolatos}, T. and {Appavuravther}, E.~Z. and {Appert}, S. and {Apple}, S.~K. and {Arai}, K. and {Araya}, A. and {Araya}, M.~C. and {Areeda}, J.~S. and {Ar{\`e}ne}, M. and {Aritomi}, N. and {Arnaud}, N. and {Arogeti}, M. and {Aronson}, S.~M. and {Arun}, K.~G. and {Asada}, H. and {Asali}, Y. and {Ashton}, G. and {Aso}, Y. and {Assiduo}, M. and {Melo}, S. Assis De Souza and {Aston}, S.~M. and {Astone}, P. and {Aubin}, F. and {Aultoneal}, K. and {Austin}, C. and {Babak}, S. and {Badaracco}, F. and {Bader}, M.~K.~M. and {Badger}, C. and {Bae}, S. and {Bae}, Y. and {Baer}, A.~M. and {Bagnasco}, S. and {Bai}, Y. and {Baird}, J. and {Bajpai}, R. and {Baka}, T. and {Ball}, M. and {Ballardin}, G. and {Ballmer}, S.~W. and {Balsamo}, A. and {Baltus}, G. and {Banagiri}, S. and {Banerjee}, B. and {Bankar}, D. and {Barayoga}, J.~C. and {Barbieri}, C. and {Barish}, B.~C. and {Barker}, D. and {Barneo}, P. and {Barone}, F. and {Barr}, B. and {Barsotti}, L. and {Barsuglia}, M. and {Barta}, D. and {Bartlett}, J. and {Barton}, M.~A. and {Bartos}, I. and {Basak}, S. and {Bassiri}, R. and {Basti}, A. and {Bawaj}, M. and {Bayley}, J.~C. and {Bazzan}, M. and {Becher}, B.~R. and {B{\'e}csy}, B. and {Bedakihale}, V.~M. and {Beirnaert}, F. and {Bejger}, M. and {Belahcene}, I. and {Benedetto}, V. and {Beniwal}, D. and {Benjamin}, M.~G. and {Bennett}, T.~F. and {Bentley}, J.~D. and {Benyaala}, M. and {Bera}, S. and {Berbel}, M. and {Bergamin}, F. and {Berger}, B.~K. and {Bernuzzi}, S. and {Bersanetti}, D. and {Bertolini}, A. and {Betzwieser}, J. and {Beveridge}, D. and {Bhandare}, R. and {Bhandari}, A.~V. and {Bhardwaj}, U. and {Bhatt}, R. and {Bhattacharjee}, D. and {Bhaumik}, S. and {Bianchi}, A. and {Bilenko}, I.~A. and {Billingsley}, G. and {Bini}, S. and {Birney}, R. and {Birnholtz}, O. and {Biscans}, S. and {Bischi}, M. and {Biscoveanu}, S. and {Bisht}, A. and {Biswas}, B. and {Bitossi}, M. and {Bizouard}, M.-A. and {Blackburn}, J.~K. and {Blair}, C.~D. and {Blair}, D.~G. and {Blair}, R.~M. and {Bobba}, F. and {Bode}, N. and {Bo{\"e}r}, M. and {Bogaert}, G. and {Boldrini}, M. and {Bolingbroke}, G.~N. and {Bonavena}, L.~D. and {Bondu}, F. and {Bonilla}, E. and {Bonnand}, R. and {Booker}, P. and {Boom}, B.~A. and {Bork}, R. and {Boschi}, V. and {Bose}, N. and {Bose}, S. and {Bossilkov}, V. and {Boudart}, V. and {Bouffanais}, Y. and {Bozzi}, A. and {Bradaschia}, C. and {Brady}, P.~R. and {Bramley}, A. and {Branch}, A. and {Branchesi}, M. and {Brau}, J.~E. and {Breschi}, M. and {Briant}, T. and {Briggs}, J.~H. and {Brillet}, A. and {Brinkmann}, M. and {Brockill}, P. and {Brooks}, A.~F. and {Brooks}, J. and {Brown}, D.~D. and {Brunett}, S. and {Bruno}, G. and {Bruntz}, R. and {Bryant}, J. and {Bucci}, F. and {Bulik}, T. and {Bulten}, H.~J. and {Buonanno}, A. and {Burtnyk}, K. and {Buscicchio}, R. and {Buskulic}, D. and {Buy}, C. and {Byer}, R.~L. and {Davies}, G.~S. Cabourn and {Cabras}, G. and {Cabrita}, R. and {Cadonati}, L. and {Caesar}, M. and {Cagnoli}, G. and {Cahillane}, C.},
        title = "{Search for continuous gravitational wave emission from the Milky Way center in O3 LIGO-Virgo data}",
      journal = {Physical Review D},
         year = 2022,
        month = aug,
       volume = {106},
       number = {4},
          eid = {042003},
        pages = {042003},
          doi = {10.1103/PhysRevD.106.042003},
archivePrefix = {arXiv},
       eprint = {2204.04523},
 primaryClass = {astro-ph.HE},
       adsurl = {https://ui.adsabs.harvard.edu/abs/2022PhRvD.106d2003A}
}

@ARTICLE{FermiLAT1,
       author = {{Ajello}, M. and {Albert}, A. and {Atwood}, W.~B. and {Barbiellini}, G. and {Bastieri}, D. and {Bechtol}, K. and {Bellazzini}, R. and {Bissaldi}, E. and {Blandford}, R.~D. and {Bloom}, E.~D. and {Bonino}, R. and {Bottacini}, E. and {Brandt}, T.~J. and {Bregeon}, J. and {Bruel}, P. and {Buehler}, R. and {Buson}, S. and {Caliandro}, G.~A. and {Cameron}, R.~A. and {Caputo}, R. and {Caragiulo}, M. and {Caraveo}, P.~A. and {Cecchi}, C. and {Chekhtman}, A. and {Chiang}, J. and {Chiaro}, G. and {Ciprini}, S. and {Cohen-Tanugi}, J. and {Cominsky}, L.~R. and {Conrad}, J. and {Cutini}, S. and {D'Ammando}, F. and {de Angelis}, A. and {de Palma}, F. and {Desiante}, R. and {Di Venere}, L. and {Drell}, P.~S. and {Favuzzi}, C. and {Ferrara}, E.~C. and {Fusco}, P. and {Gargano}, F. and {Gasparrini}, D. and {Giglietto}, N. and {Giommi}, P. and {Giordano}, F. and {Giroletti}, M. and {Glanzman}, T. and {Godfrey}, G. and {Gomez-Vargas}, G.~A. and {Grenier}, I.~A. and {Guiriec}, S. and {Gustafsson}, M. and {Harding}, A.~K. and {Hewitt}, J.~W. and {Hill}, A.~B. and {Horan}, D. and {Jogler}, T. and {J{\'o}hannesson}, G. and {Johnson}, A.~S. and {Kamae}, T. and {Karwin}, C. and {Kn{\"o}dlseder}, J. and {Kuss}, M. and {Larsson}, S. and {Latronico}, L. and {Li}, J. and {Li}, L. and {Longo}, F. and {Loparco}, F. and {Lovellette}, M.~N. and {Lubrano}, P. and {Magill}, J. and {Maldera}, S. and {Malyshev}, D. and {Manfreda}, A. and {Mayer}, M. and {Mazziotta}, M.~N. and {Michelson}, P.~F. and {Mitthumsiri}, W. and {Mizuno}, T. and {Moiseev}, A.~A. and {Monzani}, M.~E. and {Morselli}, A. and {Moskalenko}, I.~V. and {Murgia}, S. and {Nuss}, E. and {Ohno}, M. and {Ohsugi}, T. and {Omodei}, N. and {Orlando}, E. and {Ormes}, J.~F. and {Paneque}, D. and {Pesce-Rollins}, M. and {Piron}, F. and {Pivato}, G. and {Porter}, T.~A. and {Rain{\`o}}, S. and {Rando}, R. and {Razzano}, M. and {Reimer}, A. and {Reimer}, O. and {Ritz}, S. and {S{\'a}nchez-Conde}, M. and {Saz Parkinson}, P.~M. and {Sgr{\`o}}, C. and {Siskind}, E.~J. and {Smith}, D.~A. and {Spada}, F. and {Spandre}, G. and {Spinelli}, P. and {Suson}, D.~J. and {Tajima}, H. and {Takahashi}, H. and {Thayer}, J.~B. and {Torres}, D.~F. and {Tosti}, G. and {Troja}, E. and {Uchiyama}, Y. and {Vianello}, G. and {Winer}, B.~L. and {Wood}, K.~S. and {Zaharijas}, G. and {Zimmer}, S.},
        title = "{Fermi-LAT Observations of High-Energy Gamma-Ray Emission toward the Galactic Center}",
      journal = {The Astrophysical Journal},
         year = 2016,
        month = mar,
       volume = {819},
       number = {1},
          eid = {44},
        pages = {44},
          doi = {10.3847/0004-637X/819/1/44},
archivePrefix = {arXiv},
       eprint = {1511.02938},
 primaryClass = {astro-ph.HE},
       adsurl = {https://ui.adsabs.harvard.edu/abs/2016ApJ...819...44A}
}

@ARTICLE{FermiLAT2,
       author = {{Di Mauro}, Mattia},
        title = "{Characteristics of the Galactic Center excess measured with 11 years of F e r m i -LAT data}",
      journal = {Physical Review D},
         year = 2021,
        month = mar,
       volume = {103},
       number = {6},
          eid = {063029},
        pages = {063029},
          doi = {10.1103/PhysRevD.103.063029},
archivePrefix = {arXiv},
       eprint = {2101.04694},
 primaryClass = {astro-ph.HE},
       adsurl = {https://ui.adsabs.harvard.edu/abs/2021PhRvD.103f3029D}
}

@ARTICLE{WIMP2,
       author = {{Ackermann}, M. and {Ajello}, M. and {Albert}, A. and {Atwood}, W.~B. and {Baldini}, L. and {Ballet}, J. and {Barbiellini}, G. and {Bastieri}, D. and {Bellazzini}, R. and {Bissaldi}, E. and {Blandford}, R.~D. and {Bloom}, E.~D. and {Bonino}, R. and {Bottacini}, E. and {Brandt}, T.~J. and {Bregeon}, J. and {Bruel}, P. and {Buehler}, R. and {Burnett}, T.~H. and {Cameron}, R.~A. and {Caputo}, R. and {Caragiulo}, M. and {Caraveo}, P.~A. and {Cavazzuti}, E. and {Cecchi}, C. and {Charles}, E. and {Chekhtman}, A. and {Chiang}, J. and {Chiappo}, A. and {Chiaro}, G. and {Ciprini}, S. and {Conrad}, J. and {Costanza}, F. and {Cuoco}, A. and {Cutini}, S. and {D'Ammando}, F. and {de Palma}, F. and {Desiante}, R. and {Digel}, S.~W. and {Di Lalla}, N. and {Di Mauro}, M. and {Di Venere}, L. and {Drell}, P.~S. and {Favuzzi}, C. and {Fegan}, S.~J. and {Ferrara}, E.~C. and {Focke}, W.~B. and {Franckowiak}, A. and {Fukazawa}, Y. and {Funk}, S. and {Fusco}, P. and {Gargano}, F. and {Gasparrini}, D. and {Giglietto}, N. and {Giordano}, F. and {Giroletti}, M. and {Glanzman}, T. and {Gomez-Vargas}, G.~A. and {Green}, D. and {Grenier}, I.~A. and {Grove}, J.~E. and {Guillemot}, L. and {Guiriec}, S. and {Gustafsson}, M. and {Harding}, A.~K. and {Hays}, E. and {Hewitt}, J.~W. and {Horan}, D. and {Jogler}, T. and {Johnson}, A.~S. and {Kamae}, T. and {Kocevski}, D. and {Kuss}, M. and {La Mura}, G. and {Larsson}, S. and {Latronico}, L. and {Li}, J. and {Longo}, F. and {Loparco}, F. and {Lovellette}, M.~N. and {Lubrano}, P. and {Magill}, J.~D. and {Maldera}, S. and {Malyshev}, D. and {Manfreda}, A. and {Martin}, P. and {Mazziotta}, M.~N. and {Michelson}, P.~F. and {Mirabal}, N. and {Mitthumsiri}, W. and {Mizuno}, T. and {Moiseev}, A.~A. and {Monzani}, M.~E. and {Morselli}, A. and {Negro}, M. and {Nuss}, E. and {Ohsugi}, T. and {Orienti}, M. and {Orlando}, E. and {Ormes}, J.~F. and {Paneque}, D. and {Perkins}, J.~S. and {Persic}, M. and {Pesce-Rollins}, M. and {Piron}, F. and {Principe}, G. and {Rain{\`o}}, S. and {Rando}, R. and {Razzano}, M. and {Razzaque}, S. and {Reimer}, A. and {Reimer}, O. and {S{\'a}nchez-Conde}, M. and {Sgr{\`o}}, C. and {Simone}, D. and {Siskind}, E.~J. and {Spada}, F. and {Spandre}, G. and {Spinelli}, P. and {Suson}, D.~J. and {Tajima}, H. and {Tanaka}, K. and {Thayer}, J.~B. and {Tibaldo}, L. and {Torres}, D.~F. and {Troja}, E. and {Uchiyama}, Y. and {Vianello}, G. and {Wood}, K.~S. and {Wood}, M. and {Zaharijas}, G. and {Zimmer}, S. and {Fermi LAT Collaboration}},
        title = "{The Fermi Galactic Center GeV Excess and Implications for Dark Matter}",
      journal = {The Astrophysical Journal},
         year = 2017,
        month = may,
       volume = {840},
       number = {1},
          eid = {43},
        pages = {43},
          doi = {10.3847/1538-4357/aa6cab},
archivePrefix = {arXiv},
       eprint = {1704.03910},
 primaryClass = {astro-ph.HE},
       adsurl = {https://ui.adsabs.harvard.edu/abs/2017ApJ...840...43A}
}

@article{WIMP3,
  title = {Multimessenger constraints on the dark matter interpretation of the $Fermi$-LAT Galactic Center excess},
  author = {Di Mauro, Mattia and Winkler, Martin Wolfgang},
  journal = {Physical Review D},
  volume = {103},
  issue = {12},
  pages = {123005},
  numpages = {27},
  year = {2021},
  month = {Jun},
  publisher = {American Physical Society},
  doi = {10.1103/PhysRevD.103.123005},
  url = {https://link.aps.org/doi/10.1103/PhysRevD.103.123005}
}

@ARTICLE{MSGC1,
       author = {{Bartels}, Richard and {Krishnamurthy}, Suraj and {Weniger}, Christoph},
        title = "{Strong Support for the Millisecond Pulsar Origin of the Galactic Center GeV Excess}",
      journal = {Physical Review Letters},
         year = 2016,
        month = feb,
       volume = {116},
       number = {5},
          eid = {051102},
        pages = {051102},
          doi = {10.1103/PhysRevLett.116.051102},
archivePrefix = {arXiv},
       eprint = {1506.05104},
 primaryClass = {astro-ph.HE},
       adsurl = {https://ui.adsabs.harvard.edu/abs/2016PhRvL.116e1102B}
}

@ARTICLE{MSGC2,
       author = {{Lee}, Samuel K. and {Lisanti}, Mariangela and {Safdi}, Benjamin R. and {Slatyer}, Tracy R. and {Xue}, Wei},
        title = "{Evidence for Unresolved {\ensuremath{\gamma}} -Ray Point Sources in the Inner Galaxy}",
      journal = {Physical Review Letters},
         year = 2016,
        month = feb,
       volume = {116},
       number = {5},
          eid = {051103},
        pages = {051103},
          doi = {10.1103/PhysRevLett.116.051103},
archivePrefix = {arXiv},
       eprint = {1506.05124},
 primaryClass = {astro-ph.HE},
       adsurl = {https://ui.adsabs.harvard.edu/abs/2016PhRvL.116e1103L}
}

@ARTICLE{MSGC3,
       author = {{Calore}, F. and {Di Mauro}, M. and {Donato}, F. and {Hessels}, J.~W.~T. and {Weniger}, C.},
        title = "{Radio Detection Prospects for a Bulge Population of Millisecond Pulsars as Suggested by Fermi-LAT Observations of the Inner Galaxy}",
      journal = {The Astrophysical Journal},
         year = 2016,
        month = aug,
       volume = {827},
       number = {2},
          eid = {143},
        pages = {143},
          doi = {10.3847/0004-637X/827/2/143},
archivePrefix = {arXiv},
       eprint = {1512.06825},
 primaryClass = {astro-ph.HE},
       adsurl = {https://ui.adsabs.harvard.edu/abs/2016ApJ...827..143C}
}

@ARTICLE{MSGC4,
       author = {{Fermi-LAT Collaboration}},
        title = "{Characterizing the population of pulsars in the inner Galaxy with the Fermi Large Area Telescope}",
      journal = {arXiv e-prints},
         year = 2017,
        month = apr,
          eid = {arXiv:1705.00009},
        pages = {arXiv:1705.00009},
          doi = {10.48550/arXiv.1705.00009},
archivePrefix = {arXiv},
       eprint = {1705.00009},
 primaryClass = {astro-ph.HE},
       adsurl = {https://ui.adsabs.harvard.edu/abs/2017arXiv170500009F}
}

@ARTICLE{Lensing1,
       author = {{Nakamura}, T.~T. and {Deguchi}, S.},
        title = "{Wave Optics in Gravitational Lensing}",
      journal = {Progress of Theoretical Physics Supplement},
         year = 1999,
        month = jan,
       volume = {133},
        pages = {137-153},
          doi = {10.1143/PTPS.133.137},
       adsurl = {https://ui.adsabs.harvard.edu/abs/1999PThPS.133..137N}
}

@ARTICLE{Goyal2021,
       author = {{Goyal}, Srashti and {Haris}, K. and {Mehta}, Ajit Kumar and {Ajith}, Parameswaran},
        title = "{Testing the nature of gravitational-wave polarizations using strongly lensed signals}",
      journal = {Physical Review D},
         year = 2021,
        month = jan,
       volume = {103},
       number = {2},
          eid = {024038},
        pages = {024038},
          doi = {10.1103/PhysRevD.103.024038},
archivePrefix = {arXiv},
       eprint = {2008.07060},
 primaryClass = {gr-qc},
       adsurl = {https://ui.adsabs.harvard.edu/abs/2021PhRvD.103b4038G}
}

@ARTICLE{tgrlensing2,
       author = {{Liu}, Tonghua and {Biesiada}, Marek and {Tian}, Shuxun and {Liao}, Kai},
        title = "{Robust test of general relativity at the galactic scales by combining strong lensing systems and gravitational wave standard sirens}",
      journal = {Physical Review D},
         year = 2024,
        month = apr,
       volume = {109},
       number = {8},
          eid = {084074},
        pages = {084074},
          doi = {10.1103/PhysRevD.109.084074},
archivePrefix = {arXiv},
       eprint = {2404.05907},
 primaryClass = {gr-qc},
       adsurl = {https://ui.adsabs.harvard.edu/abs/2024PhRvD.109h4074L}
}

@ARTICLE{tgrlensing3,
       author = {{Fan}, Xi-Long and {Liao}, Kai and {Biesiada}, Marek and {Pi{\'o}rkowska-Kurpas}, Aleksandra and {Zhu}, Zong-Hong},
        title = "{Speed of Gravitational Waves from Strongly Lensed Gravitational Waves and Electromagnetic Signals}",
      journal = {Physical Review Letters},
         year = 2017,
        month = mar,
       volume = {118},
       number = {9},
          eid = {091102},
        pages = {091102},
          doi = {10.1103/PhysRevLett.118.091102},
       adsurl = {https://ui.adsabs.harvard.edu/abs/2017PhRvL.118i1102F}
}

@ARTICLE{tgrlensing4,
       author = {{Ezquiaga}, Jose Mar{\'\i}a and {Zumalac{\'a}rregui}, Miguel},
        title = "{Gravitational wave lensing beyond general relativity: Birefringence, echoes, and shadows}",
      journal = {Physical Review D},
         year = 2020,
        month = dec,
       volume = {102},
       number = {12},
          eid = {124048},
        pages = {124048},
          doi = {10.1103/PhysRevD.102.124048},
archivePrefix = {arXiv},
       eprint = {2009.12187},
 primaryClass = {gr-qc},
       adsurl = {https://ui.adsabs.harvard.edu/abs/2020PhRvD.102l4048E}
}

@ARTICLE{tgrlensing5,
       author = {{Baker}, Tessa and {Trodden}, Mark},
        title = "{Multimessenger time delays from lensed gravitational waves}",
      journal = {Physical Review D},
         year = 2017,
        month = mar,
       volume = {95},
       number = {6},
          eid = {063512},
        pages = {063512},
          doi = {10.1103/PhysRevD.95.063512},
archivePrefix = {arXiv},
       eprint = {1612.02004},
 primaryClass = {astro-ph.CO},
       adsurl = {https://ui.adsabs.harvard.edu/abs/2017PhRvD..95f3512B}
}

@ARTICLE{cosmography1,
       author = {{Sereno}, M. and {Jetzer}, Ph. and {Sesana}, A. and {Volonteri}, M.},
        title = "{Cosmography with strong lensing of LISA gravitational wave sources}",
      journal = {Monthly Notices of the Royal Astronomical Society},
         year = 2011,
        month = aug,
       volume = {415},
       number = {3},
        pages = {2773-2781},
          doi = {10.1111/j.1365-2966.2011.18895.x},
archivePrefix = {arXiv},
       eprint = {1104.1977},
 primaryClass = {astro-ph.CO},
       adsurl = {https://ui.adsabs.harvard.edu/abs/2011MNRAS.415.2773S}
}

@ARTICLE{cosmography2,
       author = {{Liao}, Kai and {Fan}, Xi-Long and {Ding}, Xuheng and {Biesiada}, Marek and {Zhu}, Zong-Hong},
        title = "{Precision cosmology from future lensed gravitational wave and electromagnetic signals}",
      journal = {Nature Communications},
         year = 2017,
        month = oct,
       volume = {8},
          eid = {1148},
        pages = {1148},
          doi = {10.1038/s41467-017-01152-9},
archivePrefix = {arXiv},
       eprint = {1703.04151},
 primaryClass = {astro-ph.CO},
       adsurl = {https://ui.adsabs.harvard.edu/abs/2017NatCo...8.1148L}
}

@ARTICLE{cosmography3,
       author = {{Jana}, Souvik and {Kapadia}, Shasvath J. and {Venumadhav}, Tejaswi and {Ajith}, Parameswaran},
        title = "{Cosmography Using Strongly Lensed Gravitational Waves from Binary Black Holes}",
      journal = {Physical Review Letters},
         year = 2023,
        month = jun,
       volume = {130},
       number = {26},
          eid = {261401},
        pages = {261401},
          doi = {10.1103/PhysRevLett.130.261401},
archivePrefix = {arXiv},
       eprint = {2211.12212},
 primaryClass = {astro-ph.CO},
       adsurl = {https://ui.adsabs.harvard.edu/abs/2023PhRvL.130z1401J}
}

@ARTICLE{Hannuksela:2020xor,
       author = {{Hannuksela}, Otto A. and {Collett}, Thomas E. and {{\c{C}}al{\i}{\c{s}}kan}, Mesut and {Li}, Tjonnie G.~F.},
        title = "{Localizing merging black holes with sub-arcsecond precision using gravitational-wave lensing}",
      journal = {Monthly Notices of the Royal Astronomical Society},
         year = 2020,
        month = nov,
       volume = {498},
       number = {3},
        pages = {3395-3402},
          doi = {10.1093/mnras/staa2577},
archivePrefix = {arXiv},
       eprint = {2004.13811},
 primaryClass = {astro-ph.HE},
       adsurl = {https://ui.adsabs.harvard.edu/abs/2020MNRAS.498.3395H}
}

@ARTICLE{Ng2018,
       author = {{Ng}, Ken K.~Y. and {Wong}, Kaze W.~K. and {Broadhurst}, Tom and {Li}, Tjonnie G.~F.},
        title = "{Precise LIGO lensing rate predictions for binary black holes}",
      journal = {Physical Review D},
         year = 2018,
        month = jan,
       volume = {97},
       number = {2},
          eid = {023012},
        pages = {023012},
          doi = {10.1103/PhysRevD.97.023012},
archivePrefix = {arXiv},
       eprint = {1703.06319},
 primaryClass = {astro-ph.CO},
       adsurl = {https://ui.adsabs.harvard.edu/abs/2018PhRvD..97b3012N}
}

@ARTICLE{NSlensprospects1,
       author = {{Basak}, Soummyadip and {Sharma}, Aditya Kumar and {Kapadia}, Shasvath J. and {Ajith}, Parameswaran},
        title = "{Prospects for the Observation of Continuous Gravitational Waves from Spinning Neutron Stars Lensed by the Galactic Supermassive Black Hole}",
      journal = {The Astrophysical Journal Letters},
         year = 2023,
        month = jan,
       volume = {942},
       number = {2},
          eid = {L31},
        pages = {L31},
          doi = {10.3847/2041-8213/acab50},
archivePrefix = {arXiv},
       eprint = {2205.00022},
 primaryClass = {gr-qc},
       adsurl = {https://ui.adsabs.harvard.edu/abs/2023ApJ...942L..31B}
}

@ARTICLE{NSlensprospects2,
       author = {{Savastano}, Stefano and {Vernizzi}, Filippo and {Zumalac{\'a}rregui}, Miguel},
        title = "{Through the lens of Sgr A$^{*}$: Identifying and resolving strongly lensed continuous gravitational waves beyond the Einstein radius}",
      journal = {Physical Review D},
         year = 2024,
        month = jan,
       volume = {109},
       number = {2},
          eid = {024064},
        pages = {024064},
          doi = {10.1103/PhysRevD.109.024064},
archivePrefix = {arXiv},
       eprint = {2212.14697},
 primaryClass = {gr-qc},
       adsurl = {https://ui.adsabs.harvard.edu/abs/2024PhRvD.109b4064S}
}

@ARTICLE{NSlensprospects3,
       author = {{Suyamprakasam}, Sudhagar and {Harikumar}, Sreekanth and {Cieciel{\k{a}}g}, Pawe{\l} and {Figura}, Przemys{\l}aw and {Bejger}, Micha{\l} and {Biesiada}, Marek},
        title = "{Microlensing of long-duration gravitational wave signals originating from Galactic sources}",
      journal = {Physical Review D},
         year = 2025,
        month = dec,
       volume = {112},
       number = {12},
          eid = {124019},
        pages = {124019},
          doi = {10.1103/d1ys-8fw3},
archivePrefix = {arXiv},
       eprint = {2503.21845},
 primaryClass = {gr-qc},
       adsurl = {https://ui.adsabs.harvard.edu/abs/2025PhRvD.112l4019S}
}

@ARTICLE{schodel2002star,
       author = {{Sch{\"o}del}, R. and {Ott}, T. and {Genzel}, R. and {Hofmann}, R. and {Lehnert}, M. and {Eckart}, A. and {Mouawad}, N. and {Alexander}, T. and {Reid}, M.~J. and {Lenzen}, R. and {Hartung}, M. and {Lacombe}, F. and {Rouan}, D. and {Gendron}, E. and {Rousset}, G. and {Lagrange}, A. -M. and {Brandner}, W. and {Ageorges}, N. and {Lidman}, C. and {Moorwood}, A.~F.~M. and {Spyromilio}, J. and {Hubin}, N. and {Menten}, K.~M.},
        title = "{A star in a 15.2-year orbit around the supermassive black hole at the centre of the Milky Way}",
      journal = {Nature},
         year = 2002,
        month = oct,
       volume = {419},
       number = {6908},
        pages = {694-696},
          doi = {10.1038/nature01121},
archivePrefix = {arXiv},
       eprint = {astro-ph/0210426},
 primaryClass = {astro-ph},
       adsurl = {https://ui.adsabs.harvard.edu/abs/2002Natur.419..694S}
}

@ARTICLE{ghez2003first,
       author = {{Ghez}, A.~M. and {Duch{\^e}ne}, G. and {Matthews}, K. and {Hornstein}, S.~D. and {Tanner}, A. and {Larkin}, J. and {Morris}, M. and {Becklin}, E.~E. and {Salim}, S. and {Kremenek}, T. and {Thompson}, D. and {Soifer}, B.~T. and {Neugebauer}, G. and {McLean}, I.},
        title = "{The First Measurement of Spectral Lines in a Short-Period Star Bound to the Galaxy's Central Black Hole: A Paradox of Youth}",
      journal = {The Astrophysical Journal Letters},
         year = 2003,
        month = apr,
       volume = {586},
       number = {2},
        pages = {L127-L131},
          doi = {10.1086/374804},
archivePrefix = {arXiv},
       eprint = {astro-ph/0302299},
 primaryClass = {astro-ph},
       adsurl = {https://ui.adsabs.harvard.edu/abs/2003ApJ...586L.127G}
}

@ARTICLE{gravity2019,
       author = {{GRAVITY Collaboration} and {Abuter}, R. and {Amorim}, A. and {Baub{\"o}ck}, M. and {Berger}, J.~P. and {Bonnet}, H. and {Brandner}, W. and {Cardoso}, V. and {Cl{\'e}net}, Y. and {de Zeeuw}, P.~T. and {Dexter}, J. and {Eckart}, A. and {Eisenhauer}, F. and {F{\"o}rster Schreiber}, N.~M. and {Garcia}, P. and {Gao}, F. and {Gendron}, E. and {Genzel}, R. and {Gillessen}, S. and {Habibi}, M. and {Haubois}, X. and {Henning}, T. and {Hippler}, S. and {Horrobin}, M. and {Jim{\'e}nez-Rosales}, A. and {Jochum}, L. and {Jocou}, L. and {Kaufer}, A. and {Kervella}, P. and {Lacour}, S. and {Lapeyr{\`e}re}, V. and {Le Bouquin}, J. -B. and {L{\'e}na}, P. and {Nowak}, M. and {Ott}, T. and {Paumard}, T. and {Perraut}, K. and {Perrin}, G. and {Pfuhl}, O. and {Rodr{\'\i}guez-Coira}, G. and {Shangguan}, J. and {Scheithauer}, S. and {Stadler}, J. and {Straub}, O. and {Straubmeier}, C. and {Sturm}, E. and {Tacconi}, L.~J. and {Vincent}, F. and {von Fellenberg}, S. and {Waisberg}, I. and {Widmann}, F. and {Wieprecht}, E. and {Wiezorrek}, E. and {Woillez}, J. and {Yazici}, S. and {Zins}, G.},
        title = "{Detection of the Schwarzschild precession in the orbit of the star S2 near the Galactic centre massive black hole}",
      journal = {Astronomy \& Astrophysics},
         year = 2020,
        month = apr,
       volume = {636},
          eid = {L5},
        pages = {L5},
          doi = {10.1051/0004-6361/202037813},
archivePrefix = {arXiv},
       eprint = {2004.07187},
 primaryClass = {astro-ph.GA},
       adsurl = {https://ui.adsabs.harvard.edu/abs/2020A&A...636L...5G}
}

@ARTICLE{eht2022,
       author = {{Event Horizon Telescope Collaboration} and {Akiyama}, Kazunori and {Alberdi}, Antxon and {Alef}, Walter and {Algaba}, Juan Carlos and {Anantua}, Richard and {Asada}, Keiichi and {Azulay}, Rebecca and {Bach}, Uwe and {Baczko}, Anne-Kathrin and {Ball}, David and {Balokovi{\'c}}, Mislav and {Barrett}, John and {Baub{\"o}ck}, Michi and {Benson}, Bradford A. and {Bintley}, Dan and {Blackburn}, Lindy and {Blundell}, Raymond and {Bouman}, Katherine L. and {Bower}, Geoffrey C. and {Boyce}, Hope and {Bremer}, Michael and {Brinkerink}, Christiaan D. and {Brissenden}, Roger and {Britzen}, Silke and {Broderick}, Avery E. and {Broguiere}, Dominique and {Bronzwaer}, Thomas and {Bustamante}, Sandra and {Byun}, Do-Young and {Carlstrom}, John E. and {Ceccobello}, Chiara and {Chael}, Andrew and {Chan}, Chi-kwan and {Chatterjee}, Koushik and {Chatterjee}, Shami and {Chen}, Ming-Tang and {Chen}, Yongjun and {Cheng}, Xiaopeng and {Cho}, Ilje and {Christian}, Pierre and {Conroy}, Nicholas S. and {Conway}, John E. and {Cordes}, James M. and {Crawford}, Thomas M. and {Crew}, Geoffrey B. and {Cruz-Osorio}, Alejandro and {Cui}, Yuzhu and {Davelaar}, Jordy and {De Laurentis}, Mariafelicia and {Deane}, Roger and {Dempsey}, Jessica and {Desvignes}, Gregory and {Dexter}, Jason and {Dhruv}, Vedant and {Doeleman}, Sheperd S. and {Dougal}, Sean and {Dzib}, Sergio A. and {Eatough}, Ralph P. and {Emami}, Razieh and {Falcke}, Heino and {Farah}, Joseph and {Fish}, Vincent L. and {Fomalont}, Ed and {Ford}, H. Alyson and {Fraga-Encinas}, Raquel and {Freeman}, William T. and {Friberg}, Per and {Fromm}, Christian M. and {Fuentes}, Antonio and {Galison}, Peter and {Gammie}, Charles F. and {Garc{\'\i}a}, Roberto and {Gentaz}, Olivier and {Georgiev}, Boris and {Goddi}, Ciriaco and {Gold}, Roman and {G{\'o}mez-Ruiz}, Arturo I. and {G{\'o}mez}, Jos{\'e} L. and {Gu}, Minfeng and {Gurwell}, Mark and {Hada}, Kazuhiro and {Haggard}, Daryl and {Haworth}, Kari and {Hecht}, Michael H. and {Hesper}, Ronald and {Heumann}, Dirk and {Ho}, Luis C. and {Ho}, Paul and {Honma}, Mareki and {Huang}, Chih-Wei L. and {Huang}, Lei and {Hughes}, David H. and {Ikeda}, Shiro and {Impellizzeri}, C.~M. Violette and {Inoue}, Makoto and {Issaoun}, Sara and {James}, David J. and {Jannuzi}, Buell T. and {Janssen}, Michael and {Jeter}, Britton and {Jiang}, Wu and {Jim{\'e}nez-Rosales}, Alejandra and {Johnson}, Michael D. and {Jorstad}, Svetlana and {Joshi}, Abhishek V. and {Jung}, Taehyun and {Karami}, Mansour and {Karuppusamy}, Ramesh and {Kawashima}, Tomohisa and {Keating}, Garrett K. and {Kettenis}, Mark and {Kim}, Dong-Jin and {Kim}, Jae-Young and {Kim}, Jongsoo and {Kim}, Junhan and {Kino}, Motoki and {Koay}, Jun Yi and {Kocherlakota}, Prashant and {Kofuji}, Yutaro and {Koch}, Patrick M. and {Koyama}, Shoko and {Kramer}, Carsten and {Kramer}, Michael and {Krichbaum}, Thomas P. and {Kuo}, Cheng-Yu and {La Bella}, Noemi and {Lauer}, Tod R. and {Lee}, Daeyoung and {Lee}, Sang-Sung and {Leung}, Po Kin and {Levis}, Aviad and {Li}, Zhiyuan and {Lico}, Rocco and {Lindahl}, Greg and {Lindqvist}, Michael and {Lisakov}, Mikhail and {Liu}, Jun and {Liu}, Kuo and {Liuzzo}, Elisabetta and {Lo}, Wen-Ping and {Lobanov}, Andrei P. and {Loinard}, Laurent and {Lonsdale}, Colin J. and {Lu}, Ru-Sen and {Mao}, Jirong and {Marchili}, Nicola and {Markoff}, Sera and {Marrone}, Daniel P. and {Marscher}, Alan P. and {Mart{\'\i}-Vidal}, Iv{\'a}n and {Matsushita}, Satoki and {Matthews}, Lynn D. and {Medeiros}, Lia and {Menten}, Karl M. and {Michalik}, Daniel and {Mizuno}, Izumi and {Mizuno}, Yosuke and {Moran}, James M. and {Moriyama}, Kotaro and {Moscibrodzka}, Monika and {M{\"u}ller}, Cornelia and {Mus}, Alejandro and {Musoke}, Gibwa and {Myserlis}, Ioannis and {Nadolski}, Andrew and {Nagai}, Hiroshi and {Nagar}, Neil M. and {Nakamura}, Masanori and {Narayan}, Ramesh and {Narayanan}, Gopal and {Natarajan}, Iniyan and {Nathanail}, Antonios and {Fuentes}, Santiago Navarro and {Neilsen}, Joey and {Neri}, Roberto and {Ni}, Chunchong and {Noutsos}, Aristeidis and {Nowak}, Michael A. and {Oh}, Junghwan and {Okino}, Hiroki and {Olivares}, H{\'e}ctor and {Ortiz-Le{\'o}n}, Gisela N. and {Oyama}, Tomoaki and {{\"O}zel}, Feryal and {Palumbo}, Daniel C.~M. and {Paraschos}, Georgios Filippos and {Park}, Jongho and {Parsons}, Harriet and {Patel}, Nimesh and {Pen}, Ue-Li and {Pesce}, Dominic W. and {Pi{\'e}tu}, Vincent and {Plambeck}, Richard and {PopStefanija}, Aleksandar and {Porth}, Oliver and {P{\"o}tzl}, Felix M. and {Prather}, Ben and {Preciado-L{\'o}pez}, Jorge A. and {Psaltis}, Dimitrios},
        title = "{First Sagittarius A* Event Horizon Telescope Results. I. The Shadow of the Supermassive Black Hole in the Center of the Milky Way}",
      journal = {The Astrophysical Journal Letters},
         year = 2022,
        month = may,
       volume = {930},
       number = {2},
          eid = {L12},
        pages = {L12},
          doi = {10.3847/2041-8213/ac6674},
       adsurl = {https://ui.adsabs.harvard.edu/abs/2022ApJ...930L..12E}
}

@ARTICLE{balickandbrown1974,
       author = {{Balick}, B. and {Brown}, R.~L.},
        title = "{Intense sub-arcsecond structure in the galactic center.}",
      journal = {The Astrophysical Journal},
         year = 1974,
        month = dec,
       volume = {194},
        pages = {265-270},
          doi = {10.1086/153242},
       adsurl = {https://ui.adsabs.harvard.edu/abs/1974ApJ...194..265B}
}

@article{AliciaHough,
  title = {Hough transform search for continuous gravitational waves},
  author = {Krishnan, Badri and Sintes, Alicia M. and Papa, Maria Alessandra and Schutz, Bernard F. and Frasca, Sergio and Palomba, Cristiano},
  journal = {Physical Review D},
  volume = {70},
  issue = {8},
  pages = {082001},
  numpages = {23},
  year = {2004},
  month = {Oct},
  publisher = {American Physical Society},
  doi = {10.1103/PhysRevD.70.082001},
  url = {https://link.aps.org/doi/10.1103/PhysRevD.70.082001}
}

@misc{lalsuite,
       author = {{The LIGO Scientific Collaboration}},
        title = "{LALSuite: LIGO Scientific Collaboration Algorithm Library Suite}",
 howpublished = {Astrophysics Source Code Library, record ascl:2012.021},
         year = 2020,
        month = dec,
          eid = {ascl:2012.021},
archivePrefix = {ascl},
       eprint = {2012.021},
       adsurl = {https://ui.adsabs.harvard.edu/abs/2020ascl.soft12021L}
}

@article{Fasttracks,
  title = {One-stop strategy to search for long-duration gravitational-wave signals},
  author = {Tenorio, Rodrigo and M\'erou, Joan-Ren\'e and Sintes, Alicia M.},
  journal = {Physical Review D},
  volume = {111},
  issue = {10},
  pages = {104002},
  numpages = {19},
  year = {2025},
  month = {May},
  publisher = {American Physical Society},
  doi = {10.1103/PhysRevD.111.104002},
  url = {https://link.aps.org/doi/10.1103/PhysRevD.111.104002}
}

@article{PiaHough,
  title = {Method for all-sky searches of continuous gravitational wave signals using the frequency-Hough transform},
  author = {Astone, Pia and Colla, Alberto and D'Antonio, Sabrina and Frasca, Sergio and Palomba, Cristiano},
  journal = {Physical Review D},
  volume = {90},
  issue = {4},
  pages = {042002},
  numpages = {24},
  year = {2014},
  month = {Aug},
  publisher = {American Physical Society},
  doi = {10.1103/PhysRevD.90.042002},
  url = {https://link.aps.org/doi/10.1103/PhysRevD.90.042002}
}

@ARTICLE{Leong2025,
       author = {{Leong}, Samson H.~W. and {Janquart}, Justin and {Sharma}, Aditya Kumar and {Martens}, Paul and {Ajith}, Parameswaran and {Hannuksela}, Otto A.},
        title = "{Constraining Binary Mergers in Active Galactic Nuclei Disks Using the Nonobservation of Lensed Gravitational Waves}",
      journal = {The Astrophysical Journal Letters},
         year = 2025,
        month = feb,
       volume = {979},
       number = {2},
          eid = {L27},
        pages = {L27},
          doi = {10.3847/2041-8213/ad9ead},
archivePrefix = {arXiv},
       eprint = {2408.13144},
 primaryClass = {astro-ph.HE},
       adsurl = {https://ui.adsabs.harvard.edu/abs/2025ApJ...979L..27L}
}

\end{document}